\documentclass[pra,aps,amsmath,amssymb,showpacs,floatfix,twocolumn]{revtex4-1}
\usepackage{graphicx,epsfig}
\usepackage[usenames]{color}
\usepackage{verbatim}
\newcommand{\beq}{\begin{equation}}
\newcommand{\eeq}{\end{equation}}
\newcommand{\bea}{\begin{eqnarray}}
\newcommand{\eea}{\end{eqnarray}}
\newcommand{\upa}{\uparrow}
\newcommand{\downa}{\downarrow}

\newcommand{\ket}[1]{\ensuremath{\left|#1\right\rangle}}
\newcommand{\marton}[1]{#1}
\newcommand{\martonnew}[1]{#1}
\newcommand{\br}{\mathbf{r}}

\begin{document}
\title{Confinement induced interlayer molecules: a route to strong interatomic interactions}
\author{M. Kan\'asz-Nagy$^{1}$, E. A. Demler$^2$ and G. Zar\'and$^3$}
\affiliation{$^1$Condensed Matter Research Group of the Hungarian Academy of Sciences, Budafoki \'ut 8., H-1111 Budapest, Hungary}
\affiliation{$^2$Department of Physics, Harvard University, Cambridge, MA 02138, U.S.A}
\affiliation{$^3$BME-MTA Exotic Quantum Phases Research Group, Budapest University of Technology and Economics, Budapest 1521, Hungary}

\begin{abstract}
We study theoretically the interaction between two species of ultracold atoms confined 
into two layers of a finite separation, and  demonstrate the existence of 
new types of confinement-induced interlayer \marton{bound} and quasi-bound molecules: 
these novel exciton-like interlayer molecules appear for both positive and negative 
scattering lengths, and exist even for layer separations many times larger than 
the interspecies scattering length. 
The lifetime of the quasi-bound molecules grows exponentially with increasing
layer separation, and  they can therefore be observed in simple shaking experiments,
\marton{as we demonstrate through detailed many-body calculations}.
These quasi-bound molecules can also give rise to novel interspecies Feshbach resonances, 
enabling one to control geometrically the interaction between the two species
by changing the layer separation. Rather counter-intuitively, the
species can be made strongly interacting, by increasing their spatial separation.
\martonnew{The separation induced interlayer resonances provide  a powerful tool for the experimental 
control of interspecies interactions and enables one to realize novel quantum phases of 
multicomponent quantum gases.}
\end{abstract}

\maketitle

%
%

\section{Introduction}

The  world of low-dimensional quantum systems  has attracted and continues 
to attract immense interest. In lower spatial dimensions,  interactions and quantum fluctuations both play a determining 
role, and give rise  to  exotic quantum states such as Luttinger liquids, fractional quantum Hall~\cite{fractional_q_hall} 
and quantum spin Hall states~\cite{q_spin_hall_1,q_spin_hall_2} or various kinds of spin liquid states, 
not to mention the family of high temperature superconductors, where effective two-dimensionality seems to play a 
crucial role, too~\cite{high_Tc_review}.

Ultracold atoms open radically new perspectives in studying  low-dimensional quantum systems.  
Quasi-two-dimensional and one-dimensional structures can now be created 
with ease by means of deep optical lattices~\cite{dalibard_BKT, nagerl_TG_gas}, single 'pancake' and 'cigar-shaped' traps~\cite{ketterle_pencake_trap,nagerl_cigar_shaped} or Hermite-Gaussian laser beams~\cite{hermite_gauss}, 
 and their dynamical and interaction properties can be investigated  systematically~\cite{dalibard_BKT,  nagerl_TG_gas, ketterle_pencake_trap, nagerl_cigar_shaped, chin_BKT, 
dalibard_BKT_PRL, phillips_BKT, krauth_BKT}. In fact, the rapidly improving  experimental control of these optical systems has
gradually promoted them to 'quantum simulators'~\cite{feynman_q_simulators,zwierlein__feynman_emulator}, 
and allows to test the validity of various theoretical approaches systematically. 

\begin{figure}[t]
\includegraphics[width=8.6cm]{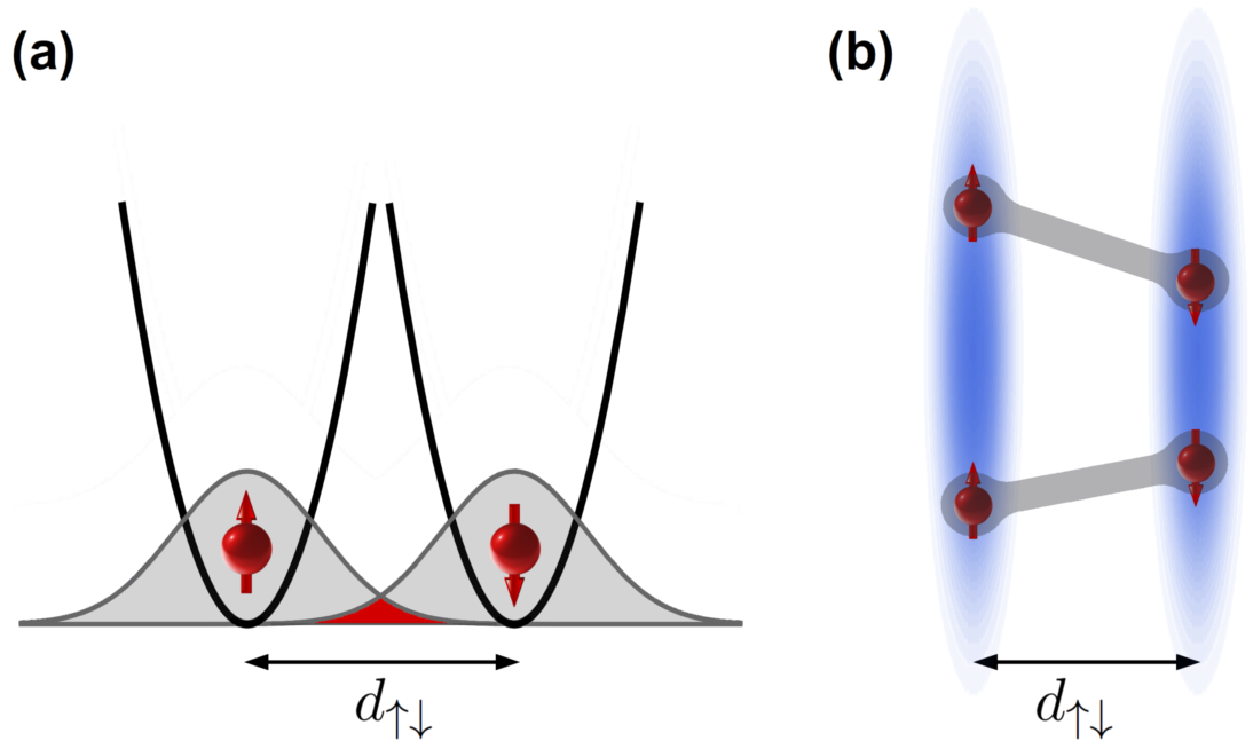}
\caption{(Color online.) 
(a) Experimental setup. A two component gas is confined into quasi-two-dimensions
using an optical lattice in the $z$ direction. \marton{
The components can be separated into 
two layers either by using a perpendicular magnetic field gradient, or through vectorial
light shifts produced by the laser field.}
(b) Schematic picture of the separated clouds.
Quasi-bound states between atoms in different layers can lead to resonant 
interlayer interactions.
}
\label{fig:ExperimentalSetup}
\end{figure}

Layered two-dimensional multi-component systems provide a particularly 
exciting perspective within the field of ultracold atoms. 
Such systems  should enable one to create the cold atomic analogues of interaction-driven
condensed matter  states such as  excitonic Bose 
condensates~\cite{Butov_Nature2002,SeamonsLilly_PRL2009} 
integer and fractional quantum-Hall
states~\cite{TutucHuse_PRL2004,KeloggEisensteinWest_2004,
EisensteinMacdonald2004,EisensteinHe1992,SuenTsui1992,LuhmanWest_PRL2008}
observed in bilayer two-dimensional quantum well structures. 
\marton{
More recently a similarly rich quantum Hall behavior~\cite{BaoPRL2010, Yacobi2009}
and the emergence of zero field magnetic phases have been observed
in bilayer graphene~\cite{WeitzYacobi2010,MayorovNovoselovScience2011,Velasco2012}.
}
As an even more interesting possibility, multicomponent fermionic and bosonic 
systems in restricted geometries  open the possibility of realizing exotic quantum states of 
matter~\cite{ketterle_zwierlein_imbalanced_SF, zwierlein_bose_fermi, ketterle_zwierlein_BEC_BCS, jin_unitarity_dynamics}, never observed before.  
\marton{
Multicomponent bosonic superfluids are expected to exhibit a rich phase diagram of 
Kosterlitz--Thouless phases, and intriguing quantum phase transitions, moreover, the
special structure of topological excitations in certain spinful condensates
is expected to modify the character of the phase transition~\cite{MoorePRL2006, Podolsky2009}.
}

To realize this rich variety of states, one obviously needs 
to \emph{understand} and  \emph{control} the interaction between  
the confined species. 
As we know from  the seminal works of Olshanii~\cite{olshanii} and 
Petrov, Holzmann, and Shlyapnikov~\cite{petrov_shlyapnikov_PRL,petrov_shlyapnikov_PRA}, confinement  
radically modifies the effective interaction between atoms~\cite{demler_pekker}; 
in quasi-two-dimensions, in particular, confinement
leads to the emergence of a bound state of energy $\epsilon = - |E^B|$ 
irrespective of the sign of the interactions of unconfined particles,
and a corresponding broad scattering resonance at 
collision energies $\epsilon \sim |E^B|$ also appears. 
Consequently, due to the presence of this bound state, the effective interaction 
of low-energy quasiparticles is always repulsive~\cite{petrov_shlyapnikov_PRL, bloch_review}.
The aforementioned bound states originate from the peculiar scattering properties of  two-dimensional systems~\cite{petrov_shlyapnikov_PRL,landau_QM}, and were indeed observed both in quasi-one and in quasi-two 
dimensions~\cite{demler_polaron_vs_kohl,nagerl_cigar_shaped, jochim_CIR, kohl_CIR, esslinger_CIR,Sommer2012}.

In this work,  we focus our attention on 
layered two-dimensional systems, sketched in Fig.~ \ref{fig:ExperimentalSetup}, and explore 
how  confinement influences interactions in such structures. To be specific, 
we consider  atoms of two hyperfine states, $\alpha = \uparrow, \downarrow$,
confined optically into quasi-two-dimensional  layers within the $xy$ plane. 
Having different magnetic moments, the separation $d_{\upa\downa}$ of the two hyperfine components
can be controlled by a magnetic field gradient applied in the $z$ direction. 
\marton{
An alternative, and maybe even simpler way to separate the layers is by using spin-dependent
optical lattices~\cite{spin_dependent_lattices1, spin_dependent_lattices2, spin_dependent_lattices3}, 
through the application of vectorial light shifts.
}

\begin{figure}[t]
\includegraphics[width=6.5cm]{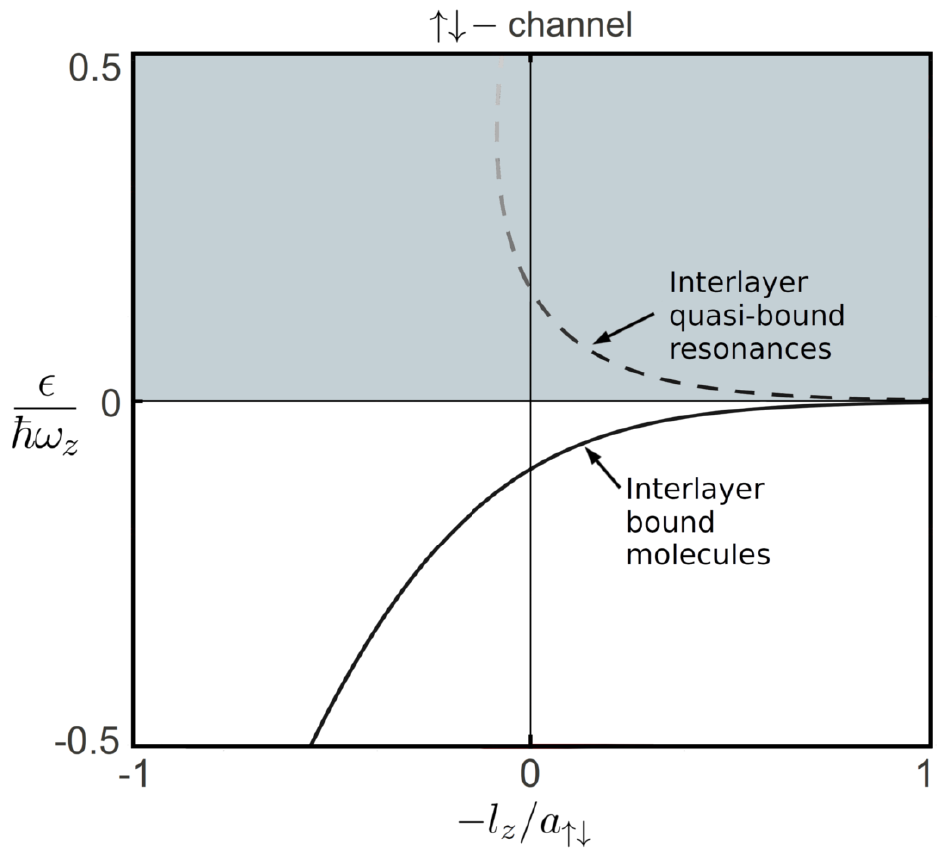}
\caption{(Color online.) Energy of the interlayer molecule in units of confinement energy, $\hbar \omega_z$, 
as a function of the oscillator length of the confining potential, $l_z$,
divided by the three-dimensional scattering length, $a_{\uparrow\downarrow}$. 
The { dashed line}  indicates the position of the corresponding resonance. 
}
\label{binding_energy_sketch}
\end{figure}

For simplicity, we assume a simple parabolic confinement for each species, 
\beq
{\cal H}_{\alpha}=
\frac{{\mathbf p}^2}{2 m} 
+ \frac{m\, \omega_z^2}{2}\, z_{\alpha}^2\;, 
\label{eq:calH} \\
\eeq
with the $z$ coordinates $z_{\alpha} \equiv z - z_{\alpha}^0$ 
measured from the centers of the layers, $z_{\alpha}^0$,
and with the natural length scale in the transverse direction set by the oscillator length 
$l_z \equiv \sqrt{\hbar / (m \omega_z)}$.
We consider a short-ranged s-wave interaction between each hyperfine components~\cite{footnote0}, 
$V_{\alpha\beta}(\br-\br')$,
characterized by the three-dimensional scattering lengths $a_{\uparrow\uparrow}$, $a_{\uparrow\downarrow}$, and  $a_{\downarrow\downarrow}$.
 As a first step, we determine the two-particle scattering wave functions and scattering amplitudes  analytically. We find that, as a result of  confinement,  
in the $\uparrow\downarrow$ channel bound interlayer molecular states of energy 
$E^B_{\upa\downa}$ emerge both for positive, $a_{\downarrow\uparrow}>0$,
and for negative  scattering length, $a_{\downarrow\uparrow}<0$   (see Fig.~\ref{binding_energy_sketch}).
As schematically shown  in Fig.~\ref{fig:ExperimentalSetup} (and later displayed in Fig.~\ref{fig:Bound_state_wave_functions}), these 'giant' molecular states extend over both layers simultaneously,
\marton{somewhat similarly to electronic excitons in bilayer quantum well structures~\cite{Butov_Nature2002,SeamonsLilly_PRL2009}}. 
Similar to the case of a single component~\cite{petrov_shlyapnikov_PRA},  their presence 
implies a \emph{repulsive} effective interaction 
between species $\uparrow$ and $\downarrow$ at low energies,
$\epsilon \ll |E^B_{\upa\downa}|$, irrespective of the 
sign of $a_{\uparrow\downarrow}$.

As another consequence of confinement, an unexpected scattering resonance 
(quasi-bound state) appears at positive collision energies 
$\epsilon \sim  |E^B_{\upa\downa}|$ (dashed line in Fig.~\ref{binding_energy_sketch})~\cite{footnote_QB_molecule}. 
\marton{Furthermore, we find similar resonances near the edges of the transverse harmonic oscillator channels,
$\nu \hbar \omega_z$, with channel index $\nu$.}
Similar \marton{quasi-bound molecular states} also exist for a single layer of atoms. There, however these confinement-induced molecular states are extremely (logarithmically) broad in energy~\cite{petrov_shlyapnikov_PRA},  and have therefore never been observed experimentally. In the layered arrangement studied here, however, the line widths of these molecular resonances are very sensitive to the layer separation, and become sufficiently sharp to be 
observable for appropriate separations. As we demonstrate through 
detailed many-body calculations  for thermal bosons,  the bound state and a
quasi-bound state  appear as separate, well-resolved lines in the 
shaking spectrum, induced  by varying   the separation 
$d_{\uparrow\downarrow}$ periodically in time. 

The molecular resonances discussed here  offer a route to control the 
interaction between different hyperfine components geometrically: 
Not only the line width, but also the energy  of the interlayer 
molecular resonance depends sensitively on layer separation. 
Eventually, the energy of the molecular resonance approaches zero 
upon changing $d_{\uparrow\downarrow}$ and -- for positive scattering 
length and tight confinement ($a_{\uparrow\downarrow}/l_z\sim 1$ )  
a sharp  \emph{interlayer Feshbach resonance} emerges as a function  of 
layer separation in the scattering amplitude of low energy particles, $\epsilon \ll \hbar\omega_z$.
Thus, rather counterintuitively, one can induce 
an extremely \emph{strong  interaction} in the $\uparrow\downarrow$ 
channel  by separating the two hyperfine species in space.
Together with confinement-tuning, -- used previously to control 
intraspecies interactions and to realize the 
Tonks--Girardeau gas~\cite{nagerl_TG_gas,WickeDruten2011,Kinoshita2004,Paredes2004}, -- 
'separation-tuning' would enable one to gain full, purely geometrical 
control of interacting, two-dimensional multicomponent systems,
and opens a route to realizing novel interaction-driven
quantum phases.

\martonnew{Separation tuning has also been predicted to lead to interaction
resonances in quasi-one-dimensional gases, exhibiting double
scattering resonances as a function of the layer separation, in case of 
positive scattering lengths~\cite{Fu_paper}. 
However, the bilayer geometry discussed here exhibits rather distinct features,
due to the peculiarities of the two-dimensional scattering, such as
the logarithmic energy broadening of scattering resonances, and the 
a finite lifetime of the associated molecular states, depending sensitively 
the layer separation.
Our work also goes beyond the few-body calculation of Ref.~\onlinecite{Fu_paper}
in that it discusses many-body aspects of these molecular states by determining
the associated peaks in a modulation spectroscopy experiment.
}

\section{Two-particle scattering} \label{sec:two-particle}

{

We start our analysis by  studying the scattering of two particles on each other
and determining  the two-particle scattering states.
Many-body effects shall be discussed 
in Sec.~\ref{sec:modulation_experiment}  in  case of a dilute Bose 
gas~\cite{footnote__fermions_vs_bosons}.

The scattering process of particles in layers $\alpha$ and $\beta$ is
governed by the Hamiltonian $\mathcal{H}_{\alpha\beta} = \mathcal{H}_\alpha + \mathcal{H}_\beta + V_{\alpha\beta}$,
and can be greatly  simplified by transforming into relative and center of mass (COM) coordinates. Defining
\begin{equation}
\mathrm{z} \, \equiv \, z_\alpha-z_\beta, \hspace{20 pt} 
\mathrm{Z} \, \equiv \, \frac{z_\alpha + z_\beta}{2}, 
\end{equation}
and likewise introducing  in plane relative ($\vec{\rho}$) and center of mass coordinates (${\mathbf{R}}$),  
  the center of mass and relative motions decouple completely for the parabolic confinement considered,  
and  the Hamiltonian can be divided into relative (rel) and center of mass (COM) parts as
\bea
\mathcal{H}_\mathrm{rel} &=& \frac{\mathbf{p}^2_{\rho} + p_\mathrm{z}^2}{m}
\, + \, \frac{m \omega_z^2}{4} \, \mathrm{z}^2  \, + \, V_{\alpha\beta}(\vec{\rho},\,\mathrm{z} - d_{\alpha\beta}), \nonumber 
\\
\mathcal{H}_\mathrm{COM} &=& \frac{\mathbf{p}^2_{\mathbf{R}} + p_\mathrm{Z}^2}{4 m}
\, + \, m \omega_z^2 \, \mathrm{Z}^2, \nonumber
\eea
where $d_{\alpha\beta}$ denotes the separation between the layers of atoms $\beta$ and $\alpha$~\marton{\cite{footnote_Hrel}}.
All  non-trivial physics  is now contained  in the relative motion of the particles, governed  by $\mathcal{H}_\mathrm{rel}$,  
which describes the motion of a particle of reduced mass $m/2$, 
confined into quasi-two dimensions by a parabolic potential, and scattered by the
interaction potential, as shown in Fig.~\ref{fig:InteractionPotential}(a).  Notice that for the $\uparrow\uparrow$ and $\downarrow\downarrow$ 
channels the delta potential  induced by the atom-atom interaction is at the minimum of the parabolic confinement, 
while in the  $\uparrow\downarrow$ channel it is shifted to $\mathrm{z} = d_{\uparrow\downarrow}$.

\begin{figure}[t]
\includegraphics[width=8.6cm]{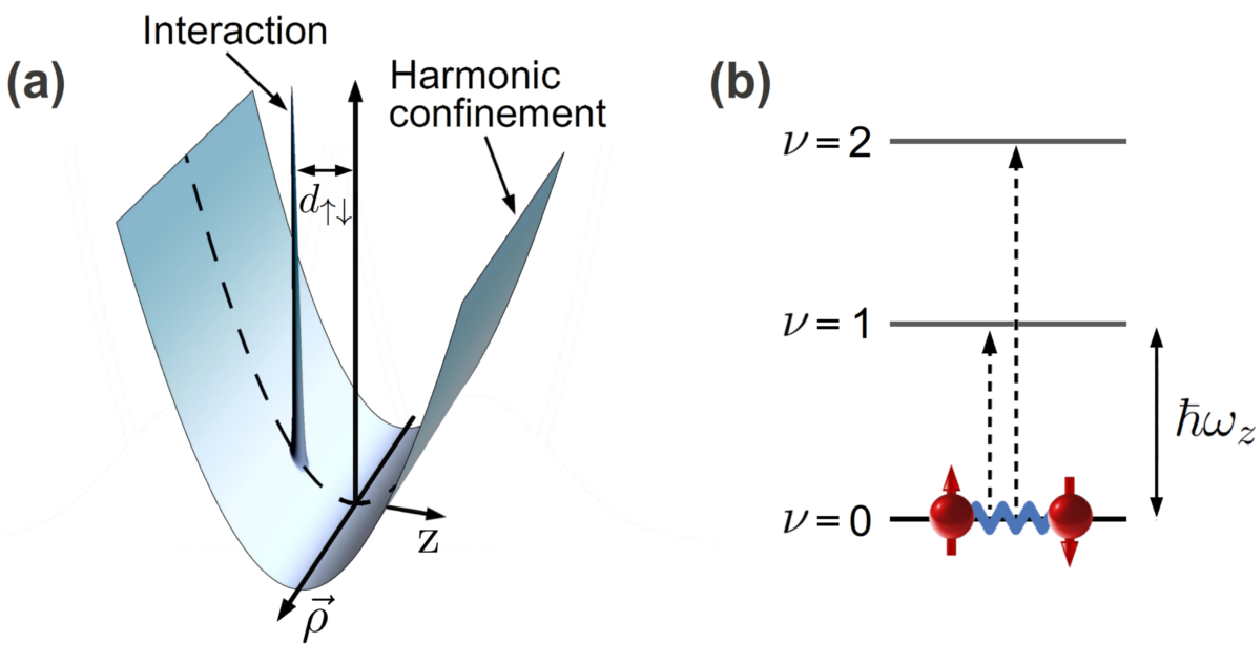}
\caption{(Color online.) 
(a) Interaction and harmonic trapping potential in relative coordinates of two particles
of opposite spins.
(b) The interaction potential mixes the relative harmonic oscillator quantum channels, $\nu$,
by introducing real and virtual transitions between them.
These  virtual transitions induce  scattering resonances at energies  close to 
the  harmonic oscillator thresholds, $\hbar\,\nu\,\omega_z$.
}
\label{fig:InteractionPotential}
\end{figure}

In the absence of interactions, 
the two particles' wave function in the $z$ direction can be expressed as 
$\Psi\sim \varphi_N(\sqrt{2}\mathrm{Z}) \varphi_\nu(\mathrm{z}/\sqrt{2})$, with 
$\varphi_\nu$  denoting the usual harmonic oscillator wave functions, and $N$ and $\nu$  
the center of mass and relative motions' quantum numbers, respectively. 
Even in the presence of interactions,  the center of mass motion remains trivial,
and the two-particle eigenstates can still be decomposed as
\beq
\tilde{\Psi} \sim {\varphi}_N(\sqrt{2}\mathrm{Z}) \, e^{i\mathbf{Q R}} \cdot \Psi_\mathrm{rel}(\vec{\rho}, \, \mathrm{z}),
\nonumber
\eeq
where  $\mathbf{Q} $ denotes the  total momentum of the particles  and
$\Psi_\mathrm{rel}$ stands for the non-trivial relative part of the wave function, governed by $\mathcal{H}_\mathrm{rel}$~\cite{footnote__symmetry_for_bosons_and_fermions}.
The total energy of the two particles can therefore  be expressed as a sum of the energy of the 
relative motion, $\epsilon$, and that of the COM motion,   
$$E = \epsilon +  E_\mathrm{COM},$$
 with 
$E_\mathrm{COM} = \mathbf{Q}^2/4m + N \hbar\omega_z\;$.

\subsection{Scattering states}

In the rest of this section, we shall  focus only on  the non-trivial relative motion,  and consider the scattering of two incoming particles 
with opposite momenta $\pm \mathbf{q}$ in the   relative  harmonic oscillator channel $\nu$.
This pair of particles   can scatter into channel $\nu'$ provided that the outgoing channel is 'open', i.e.
$\nu' \; \hbar \omega_z$ is less than the  energy of  the  
relative motion, $\epsilon = \hbar^2\mathbf{q}^2/m + \nu \hbar\omega_z$.

The corresponding two-dimensional scattering processes are characterized by the dimensionless
scattering amplitudes, $f_{\alpha\beta}^{\nu\nu'}$,
governing the long  distance ($\rho \gg l_z$)
behavior of the scattering eigenstates $\Psi_{\alpha\beta}^{\nu,\epsilon}$ 
of the relative Hamiltonian~\cite{petrov_shlyapnikov_PRA}, 
\beq
\Psi_{\alpha\beta}^{\nu,\epsilon}(\mathbf{r}) \approx \phi_\nu(\mathrm{z}) e^{i \mathbf{q} \vec{\rho}} - 
\sum_{\nu'} f_{\alpha\beta}^{\nu\nu'}(\epsilon)
\sqrt{\frac{i}{8\pi q_{\nu'} \rho}}\, e^{i q_{\nu'} \rho} \phi_{\nu'}(\mathrm{z}),
\label{eq:scattering_amplitudes}
\eeq
where $q_{\nu'} = \sqrt{m(\epsilon - \nu' \hbar\omega_z + i 0^+)}/\hbar$ 
denotes the momenta in the outgoing channels and  $\phi_\nu$ stands for the properly normalized relative wave function, $\phi_\nu(z)= \varphi_\nu(z/\sqrt 2)/2^{1/4} $~\marton{\cite{footnote2}}. 
The scattering amplitude $f_{\alpha\beta}^{\nu\nu'}$ is related to  the 
scattering  cross-section  of  $\nu\to\nu'$ 
transitions~\cite{footnote__factor_of_two}
\beq
\sigma_{\alpha\ne \beta}^{\nu\to\nu'} (q) = \frac{|f_{\alpha\beta}^{\nu\nu'}(\hbar^2 q^2 /m)|^2}{4 \;q}
\label{eq:cross_section}
\eeq
and, being dimensionless, it only depends on the three
dimensionless variables $\epsilon / \hbar\omega_z$, $d_{\alpha\beta}/l_z$ and $a_{\alpha\beta} / l_z$.

To determine the  amplitudes $f_{\alpha\beta}^{\nu\nu'}$, we  construct the two-particle scattering 
states.   We first notice that  being a scattering state, $\Psi_{\alpha\beta}^{\nu,\epsilon}$  satisfy 
the Lippmann--Schwinger equation~\cite{messiah}, 
\beq
\Psi_{\alpha\beta}^{\nu,\epsilon}(\mathbf{r}) = \phi_\nu(\mathrm{z})e^{i\mathbf{q} \vec{\mathbf{\rho}}} 
- \frac{m}{\hbar^2} \int d^3r'\, G^{(0)}_\epsilon(\mathbf{r},\mathbf{r'}) V_{\alpha\beta}(\mathbf{r}')\Psi_{\alpha\beta}^{\nu,\epsilon}(\mathbf{r}').
\label{eq:self-consistency}
\eeq
Here $\mathbf{r} = (\vec{\rho},\mathrm{z})$, and  $G^{(0)}_\epsilon$ denotes the retarded Green's function of  the non-interacting  confined system, satisfying
 $(\epsilon - {\cal H}^0_{\rm rel}) \,G^{(0)}_\epsilon(\br,\br') = \delta(\br-\br') $, and expressed in terms of modified Bessel functions as 
\beq
G^{(0)}_\epsilon(\mathbf{r},\mathbf{r}') = 
 \frac1{2\pi}\sum_{\nu=0}^\infty  {\phi_\nu(\mathrm{z}) \phi_\nu(\mathrm{z}')} K_0(-i q_\nu |\vec{\rho}-\vec{\rho}'|).
\label{eq:retarded_propagator}
\eeq

The  second term of  Eq.~\eqref{eq:self-consistency} describes the scattered part of the wave function, $\delta \Psi_{\alpha\beta}^{\nu,\epsilon}$. 
For the short-ranged potential considered here Eq.~\eqref{eq:self-consistency} immediately yields 
 \beq
 \delta \Psi_{\alpha\beta}^{\nu,\epsilon} = A \, 
 G^{(0)}_\epsilon(\mathbf{r},d_{\alpha\beta}\,\hat{\mathbf{z}}).
 \label{eq:scattered}
 \eeq
%
The  value of the  unknown proportionality constant (i.e., the amplitude of the scattered wave)
can be determined by inspecting the wave function 
around the point of interaction  at  short distances, 
$\delta r \equiv  |\mathbf{r}-d_{\alpha\beta}\, \hat{\mathbf{z}}| \ll l_z$.
At such short distances the propagation of the particles is essentially free, 
and correspondingly, $G^{(0)}_\epsilon$  exhibits the well-known $1/\delta r$ singularity of the three-dimensional free propagator,
\beq
G^{(0)}_\epsilon(\mathbf{r},d_{\alpha\beta}\hat{\mathbf{z}}) \approx \frac{1}{4 \pi} \left( \frac{1}{\delta r} 
+ \frac{w_{\alpha\beta}(\epsilon/\hbar\omega_z)}{\sqrt{2\pi}l_z} + \hdots \right).
\label{eq:G_short_distance}
\eeq
Here   $w_{\alpha\beta}(\epsilon/\hbar\omega_z)$   are energy (and separation)  dependent constants~\cite{petrov_shlyapnikov_PRA},
incorporating the effects of confinement. They can be expressed from Eq.~\eqref{eq:retarded_propagator} by carefully separating the 
$1/\delta r$ singularity (see Appendix~\ref{app:Greens_function})~\cite{petrov_shlyapnikov_PRA,demler_pekker},
 yielding
\beq
w_{\alpha\beta} ( x) = \lim_{\overline{\nu}\rightarrow \infty}
\left[ c_{\bar \nu} \, - \sum_{\nu=0}^{2 \bar \nu - 1} 
\frac{{\phi}^2_{\nu}( d_{\alpha\beta})}{{\phi}^2_{0}(0)}
  \log{\left( \frac{ \nu-x - i 0^+}{2}
   \right)}\right],
\label{eq:w_explicit}
\eeq
with  $c_\nu \equiv 2\sqrt{\frac{\nu}{\pi}}\log{\frac{\nu}{e^2}}$. 

The amplitude of the  scattered part of the wave function in Eq.~\eqref{eq:scattered}
can now be determined from the observation~\cite{petrov_shlyapnikov_PRA} 
that at short distances,  $\delta r\ll l_z$, confinement does not modify the interactions, and therefore, beyond the range of the inter-particle interaction
 -- only a few Bohr radius in practice --  the relative wave function of the two particles 
must have the same asymptotics as in three dimensions, 
\beq
\Psi_{\alpha\beta}(\mathbf{r}) \sim  1- \frac {a_{\alpha\beta}}{\delta r} +{\cal O}(\delta r)\;.
\eeq
Comparing this expansion to the asymptotic form \eqref{eq:G_short_distance} and to Eq.~\eqref{eq:scattered}, we can determine the unknown 
amplitude in Eq.~\eqref{eq:scattered}, and 
obtain the exact  solution of the two-particle scattering problem~\cite{footnote__exact_scatt_state},
\beq
\Psi_{\alpha\beta}^{\nu,\epsilon}(\mathbf{r}) = e^{i \mathbf{q}\vec{\mathbf{\rho}}} \phi_\nu(\mathrm{z})
- \frac{4 \pi \, a_{\alpha\beta} \, \phi_\nu(d_{\alpha\beta})}
{1 + \frac{a_{\alpha\beta}}{\sqrt{2\pi} \, l_z} \, w_{\alpha\beta}(\epsilon/\hbar\omega_z)} \,
G^{(0)}_{\epsilon}(\mathbf{r},d_{\alpha\beta}).
\label{eq:scattering_states_exact}
\eeq

\subsection{Scattering amplitudes, bound molecular states}
The scattering wave function,  Eq.~\eqref{eq:scattering_states_exact}, contains a lot of information. 
First,  it allows us to determine  the 
\emph{scattering amplitudes}  by comparing the asymptotic form of  $G^{(0)}_{\epsilon}(\br)$ in \eqref{eq:scattering_states_exact}
to the usual expansion of the scattering states, Eq.~\eqref{eq:scattering_amplitudes}. This yields the 
 quasi-two-dimensional scattering amplitudes
\beq
f_{\alpha\beta}^{\nu\nu'}(\epsilon) = 
\frac{ 4 \pi a_{\alpha\beta}\; 
   \phi_{\nu}( d_{\alpha\beta}) \phi_{\nu'}(d_{\alpha\beta})
  }{
  1 +    \frac {a_{\alpha_\beta}}{\sqrt {2\pi} \; l_z}  
  \,  w_{\alpha\beta}(\epsilon/\hbar\omega_z)}
 \label{eq:T-matrix_2body}
\eeq
in the open channels. The numerator of this expression conforms to the 
expectation that, to leading order, the scattering amplitude
should be proportional to the first order matrix element of the (renormalized) interaction 
with the harmonic oscillator eigenstates of the channels involved.

This naive result is, however, modified by virtual transitions  between transverse channels,
described by the functions $w_{\alpha\beta}$ in the denominator of \eqref{eq:T-matrix_2body}. 
  These functions, given by Eq.~\eqref{eq:w_explicit},
determine the positions of bound states and resonances in the presence of confinement:  these latter
emerge,  whenever the real part of the denominator in Eq.~\eqref{eq:T-matrix_2body} becomes zero.
While for scattering  within the same spin channels  we have
 $d_{\uparrow\uparrow}=d_{\downarrow\downarrow}=0$, and Eq.~\eqref{eq:T-matrix_2body}
reduces to the expression of Ref.~\onlinecite{petrov_shlyapnikov_PRA},  
for the spin $\uparrow\downarrow$ channel $f_{\uparrow\downarrow}(\epsilon)$
depends sensitively on the distance $d_{\upa\downa}$ 
between the two layers through the relative wave functions, 
$\phi_\nu$, appearing in Eq.~\eqref{eq:w_explicit}.

As a peculiar  feature of  quasi-two-dimensional scattering, 
the scattering amplitudes always have a pole at some $\epsilon \equiv E_{\alpha\beta}^B < 0$ corresponding 
to a \emph{bound molecular state},  for \emph{any} value and sign of the three-dimensional scattering length. 
Mathematically, this follows from the logarithmic singularity of $w_{\alpha\beta}$ 
at small energies, 
$w_{\alpha\beta}  \sim  \phi_0^2(d_{\alpha\beta})\,\left(-\ln\left|\epsilon/(\hbar\omega_z)\right|+i\Theta(\epsilon)\right)$, 
 related to two-dimensional propagation.
The presence of this logarithmic  singularity necessarily implies the emergence of a bound state (pole in $f^{00}_{\alpha\beta}$).   
 Specifically, for small negative scattering lengths the molecular bound state is located approximately at 
 \beq
 E^{B}_{\alpha\beta} \propto
 - \hbar\, \omega_z\; e^{-1/ \left(|a_{\alpha\beta}| {\phi^2_0(d_{\alpha\beta})}\right)}.
 \eeq   
Its wave  function  can  easily be obtained from Eq.~\eqref{eq:self-consistency},
leading to the simple form 
$$
\Psi_{\alpha\beta}^B(\mathbf{r}) = G^{(0)}_{E_{\alpha\beta}^B}(\mathbf{r},d_{\alpha\beta})\;.
$$
The interlayer molecular states  are visualized in Fig.~\ref{fig:Bound_state_wave_functions},
showing the bound state wave function in relative coordinates, as well as their
density in the laboratory frame. They  clearly display a double-peak 
feature at non-zero separation, consistent with naive expectations.
 We emphasize again that the appearance of these bound states for  negative three-dimensional scattering lengths  is the special feature
  of two-dimensional scattering~\cite{petrov_shlyapnikov_PRA}.

\begin{figure}[t]
\includegraphics[width=8.6cm]{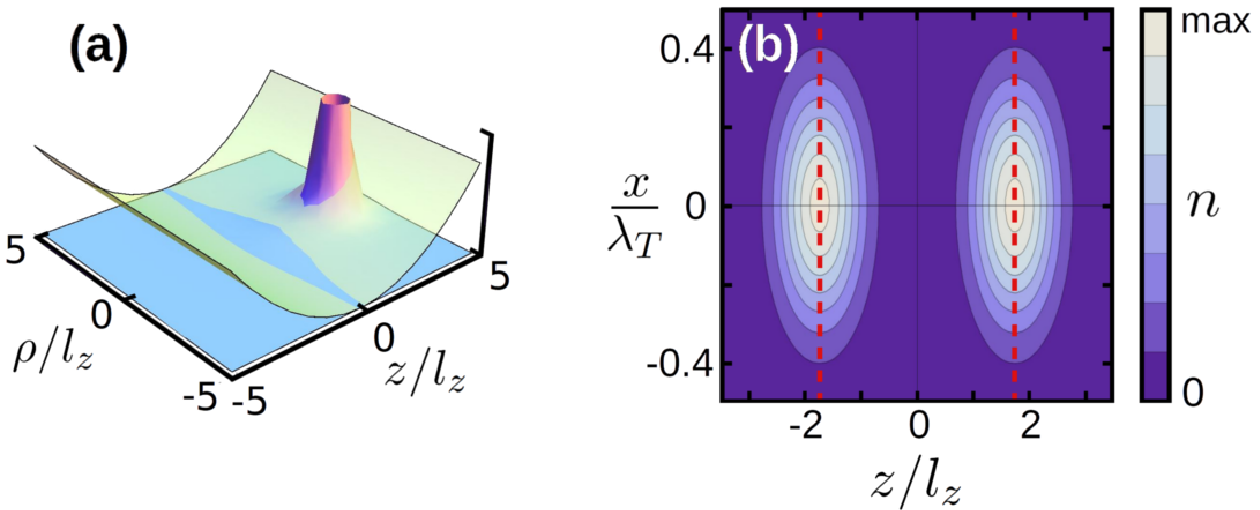}
\caption{(Color online.) Visualization of an intralayer molecular state,
at a separation $d_{\upa\downa}/l_z=3.5$.
(a) Bound state wave function $\Psi_{\upa\downa}^B$, in relative coordinates of the atoms,
displaying the well-known $1/(4 \pi \, \delta r)$ singularity of unconfined bound states at
the point of interaction. The harmonic confining potential is also shown.
(b) Real space density 
$n(\mathbf{r}) = \int d^3 \mathbf{r}' \, \| \tilde{\Psi}_{\alpha\beta} (\mathbf{r},\mathbf{r'}) \|^2$ 
of the two-particle bound state 
$\tilde{\Psi}_{\alpha\beta} (\mathbf{r},\mathbf{r'})$ in 
the $(x,z)$ plane of the laboratory frame.
Only a tiny part of the wave function
tunnels into  the intermediate region.
For the COM part of the wave function we assumed a temperature $k_B T = 0.1 \, \hbar\omega_z$
and a Gaussian in plane wave function localized within the thermal de Broglie wavelength.
 %
}
\label{fig:Bound_state_wave_functions}
\end{figure}

\subsection{Quasi-bound molecular states \label{subsec:QB_states}}

For energies $\epsilon>0$ no bound state can exist since any molecule can decay into the two-particle continuum, 
leading to a finite  imaginary part of the scattering amplitudes.  Nevertheless, the scattering of atoms still becomes structured 
due to confinement, and exhibits resonances~\cite{petrov_shlyapnikov_PRA}.
The low energy scattering amplitude, e.g.,  can  be expressed in terms of the bound state's energy as
\beq
f^{00}_{\alpha\beta} (\epsilon\approx 0) \approx \frac{4 \pi}{\log\bigl |  {E^B_{\alpha\beta}}/{\epsilon} \bigr| + i \,\pi\, \Theta(\epsilon)},
\label{eq:f_2D}
\eeq
and exhibits  a very broad resonance  at an  energy $\epsilon = |E^B_{\alpha\beta}|$~\cite{petrov_shlyapnikov_PRA}.
Similarly,  $f_{\alpha\beta}(\epsilon)$ displays resonances of finite 
width each time the real part of the denominator 
of Eq.~\eqref{eq:T-matrix_2body} crosses zero, signifying \emph{quasi-bound states} of finite lifetime.
These resonances correspond the \emph{unstable} molecular states, which then decay into the continuum.

The energies of these quasi-bound molecular states are displayed in Fig.~\ref{fig:QB} 
for some typical confinement parameters as a function of   $l_z/a_{\alpha\beta}$.  
Surprisingly, the interlayer scattering (solid line) displays features completely  absent 
in intralayer scattering (dashed lines). While for intralayer scattering 
$\nu=0\to\nu =1$ relative quantum number transitions are forbidden by 
reflection symmetry (as well as by Bose statistics in case of colliding bosons), 
such interlayer processes are allowed once 
$d_{\uparrow\downarrow} \ne0$, and they amount in the emergence of a  novel quasi-bound molecular 
state (resonance) at an energy 
\beq
   (E^{1}_{\upa\downa} - \hbar \,\omega_z)\; \propto \;
 -  \hbar \,\omega_z\, e^{-1/\left(|a_{\upa\downa}| {\phi^2_1(d_{\upa\downa})}\right)}\;.
 \eeq   
Importantly, while the weight and the binding energy of this molecular resonance
is determined by $\phi^2_1(d_{\upa\downa})$, its decay rate  
is proportional  to  $\phi^2_0(d_{\upa\downa})$,
\beq
\marton{
f^{00}_{\upa\downa} (\epsilon \lesssim \hbar\omega_z) \approx 
\frac{4 \pi}{ 
\frac{\phi_1^2(d_{\alpha\beta})}{\phi_0^2(d_{\alpha\beta})} \cdot
\log\bigl |  {E^1_{\upa\downa}}/{\epsilon} \bigr| \, + \, i \,\pi}.
}
\eeq
 Therefore, increasing the separation between the two  layers of atoms,
one can  make the quasi-bound state sharper and sharper --- 
at the cost of somewhat decreasing its weight 
(see also the inset of Fig.~\ref{fig:QB}). Similar interlayer 
quasi-bound states  of energy $E_{\upa\downa}^\nu$ appear close to every threshold, 
$\epsilon\approx  \nu \,\hbar\,\omega_z$, and can turn to a narrow resonance as one increases further 
the layer separation $d_{\upa\downa}$.

We should emphasize that Fig.~\ref{fig:QB} displays only the relative energy $\epsilon$
of the molecular states in the center of mass frame. 
The total energy of two particles is, however, 
given as a sum of the energy associated with their relative
and center of mass motions, $E= \epsilon + E_{\rm COM} $.
Accordingly, the bound state spectrum in Fig.~\ref{fig:QB} is replicated 
at energies $E\to \epsilon + N \hbar \omega_z$,   corresponding to excited
molecular bound states with an oscillating center of mass motion along 
the $z$ direction.   
One can thus observe  molecular bound states even at positive total energies $E$
in the $N>0$ channels, as long as the COM and relative motions are completely decoupled~\cite{footnote3}.

}

\begin{figure}[b]
\includegraphics[width=8.6cm]{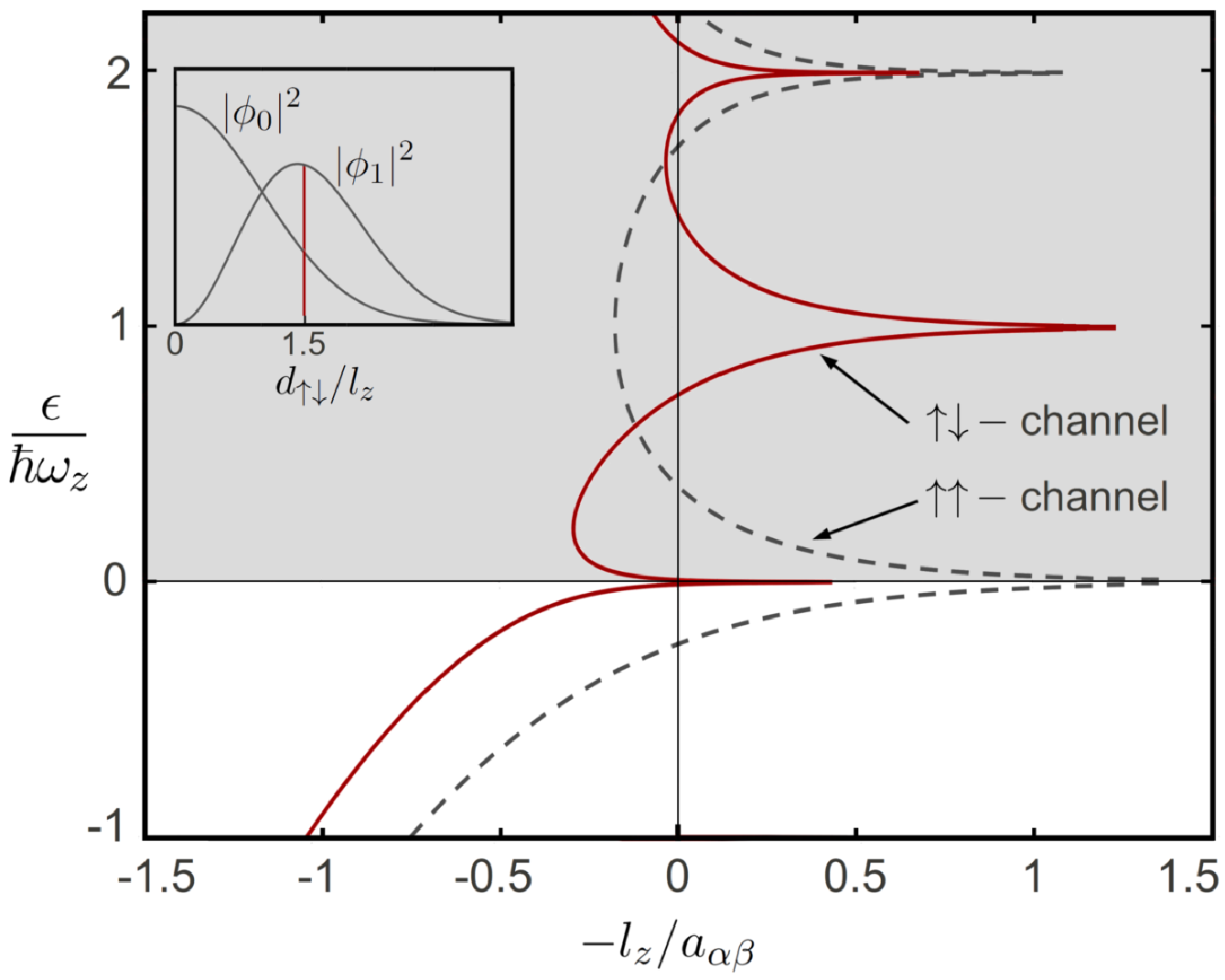}
\caption{(Color online.) 
{
Energies of bound and quasi-bound molecules in the $\upa\upa$ and $\downa\downa$ (dashed line),
and in the $\upa\downa$-channel with layer separation $d_{\upa\downa}/l_z=1.5$ (full line),
in the vicinity of a three-dimensional Fechbach resonance of the scattering length $a_{\alpha\beta}$.
Only the energy associated with the relative motion are shown. 
Inset: amplitudes $\phi_{0}^2(d_{\upa\downa})$ and $\phi_{1}^2(d_{\upa\downa})$ as functions of
$d_{\upa\downa}/l_z$. The factor $\phi_{1}^2(d_{\upa\downa})$ determines the binding energy of the first quasi-bound molecule, 
while $\phi_{0}^2(d_{\upa\downa})$ is proportional to its lifetime. }
} 
\label{fig:QB}
\end{figure}

\begin{figure}[t]
\includegraphics[width=8.6cm]{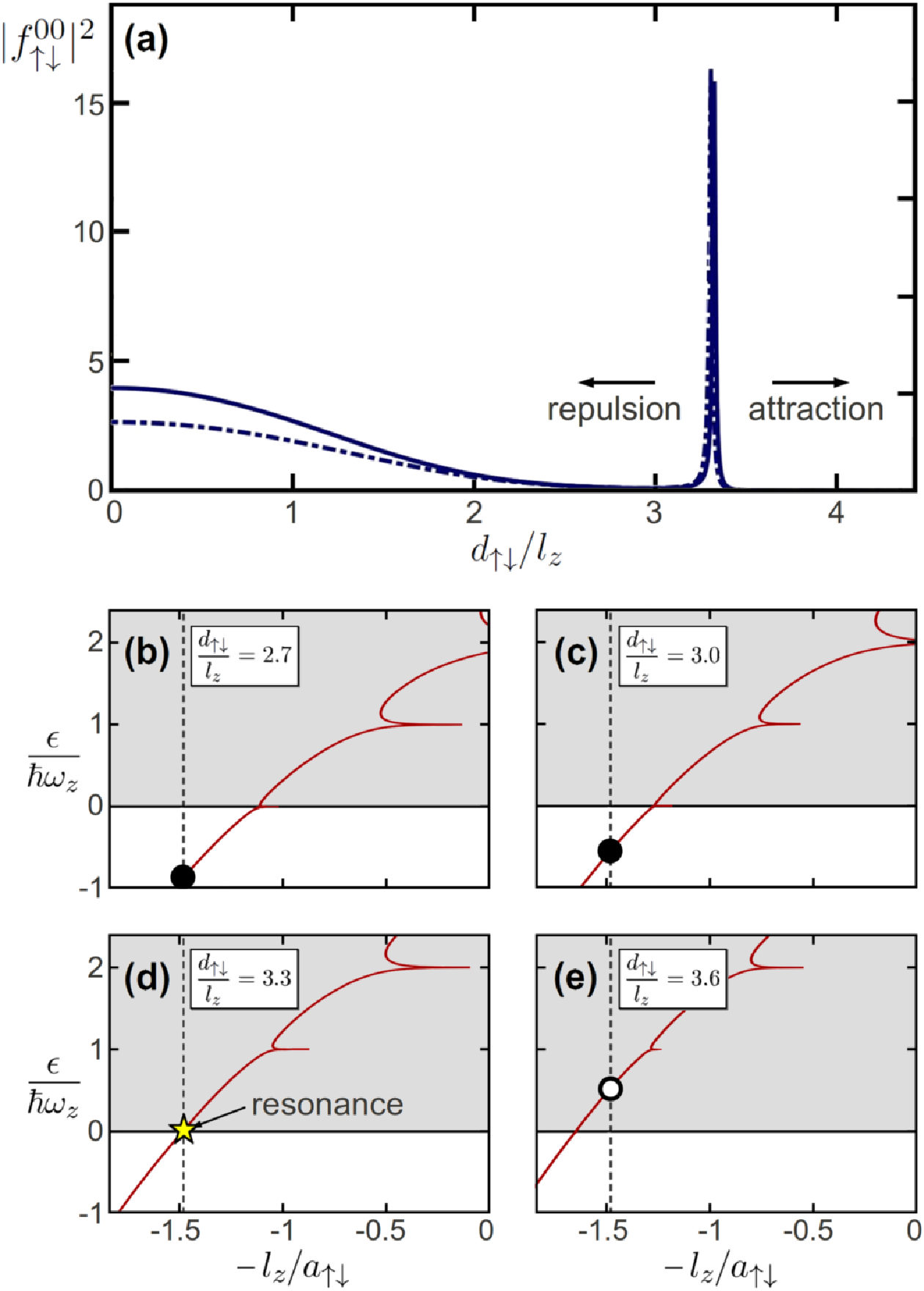}
\caption{(Color online.) 
(a) Scattering amplitude $f_{\upa\downa}^{00}$ as a function of the layer separation, $d_{\upa\downa}$,
at a fixed scattering length $a_{\uparrow\downarrow}=0.68 \, l_z$ (indicated by dashed lines in (b)-(e)). 
The continuous and  dashed curves correspond to energies $\epsilon/\hbar\omega_z=0.05$ and  $0.01$, 
respectively. A sharp Feshbach resonance structure emerges at $d_{\upa\downa}/l_z = 3.3$,
when the energy of incoming particles become resonant with a long-lived quasi-bound molecular state.
(b)-(e) The energy of the bound and quasi-bound states (full and open circles, respectively) 
at increasing layer separations 
$d_{\upa\downa} / l_z = 2.7$, $3.0$, $3.3$ and $3.6$.
The interaction resonance in (a) corresponds to the appearance of a quasi-bound molecular state at
zero energy, shown in (d). The energy of this state gets shifted to positive energies at 
larger separations, as depicted in (e).
} 
\label{fig:f_updown}
\end{figure}

\begin{figure}[t]
\includegraphics[width=8.6cm]{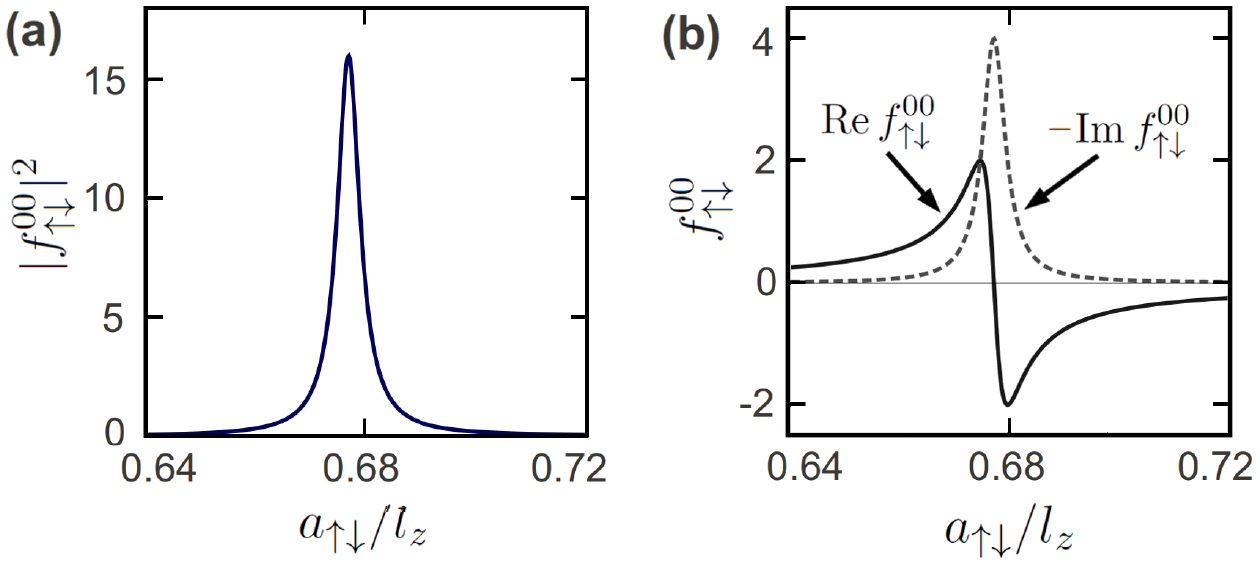}
\caption{(Color online.) Scattering amplitude $f_{\upa\downa}^{00}$ in terms of 
the scattering length $a_{\upa\downa}$ in the vicinity of the
resonance shown in Fig.~\ref{fig:f_updown}(a), but with the layer separation fixed at $d_{\upa\downa}=3.3 l_z$.
(a) The scattering amplitude exhibits a strong peak as $a_{\upa\downa}$ is 
tuned through the Feshbach-like resonance at $a_{\upa\downa} \approx 0.68 \, l_z$.
(b) Crossing the resonance, interactions turn from repulsive to attractive, 
exhibiting a large imaginary part, due to the finite lifetime of the 
resonant quasi-bound molecule.
}
\label{fig:f_updown_B}
\end{figure}

%
%

\section{Geometric interaction control \label{sec:interaction_control}}

In this section, we discuss in the case of a degenerate Bose gas, how the 
emergent interlayer resonances can be exploited to tune interspecies interactions 
\emph{independently}, simply by changing the layer separation.
In a strongly confined $(k_B T \ll \hbar \omega_z)$ gas, 
the effective interaction is approximately proportional to the scattering amplitude at the
corresponding energy $\epsilon = 2\mu$,
\beq
g_{\upa\downa} \simeq \frac{\hbar^2}{m} f_{\upa\downa}^{00}(2\mu),
\nonumber
\eeq
with the chemical potential $\mu$~\cite{petrov_shlyapnikov_PRA, demler_pekker}.
Fig.~\ref{fig:f_updown}(a) shows the 
scattering amplitude in the $\upa\downa$-channel as a function of  layer separation 
for fixed energies, $0 < \epsilon \ll \omega_z$.
As one would naively expect, for the parameters chosen, the interaction initially decreases 
with increasing separation, due to the ever weaker overlap between the atomic clouds of the layers.
Then, a sharp Feshbach-resonance-like structure emerges as a
quasi-bound state approaches the energy of the incoming particles, leading to a very strong 
interaction between the two species. 

We find similar Feshbach-like resonances at fixed layer separations, shown in Fig.~\ref{fig:f_updown_B},
as the three-dimensional scattering amplitude $a_{\upa\downa}$ is varied 
through the confinement-induced resonance.
Crossing the resonance, the effective interaction turns from repulsive to attractive, reaching its
universal, purely imaginary value of $f_{\upa\downa}^{00} = -4 i$ on resonance.

Notice, that in contrast to single layer systems~\cite{petrov_shlyapnikov_PRA}, 
these resonances appear \emph{both} on the attractive and on the repulsive 
side of the three-dimensional Feshbach resonance,
and, in a somewhat unusual way, they become the sharpest on the repulsive side, $a_{\upa\downa} > 0$.
In addition, increasing layer separation leads to the emergence of quasi-bound molecular states 
at smaller and smaller values of the scattering lengths, $a_{\upa\downa}>0$,
as indicated in Figs.~\ref{fig:QB}(b)-(e). 
The appearance of these states leads to confinement-induced resonances
also at relatively small values of the scattering lengths, $a_{\upa\downa} \lesssim l_z$.
Thus, geometrical interaction control
shall be useful for reaching the strongly correlated regime
in systems, where no magnetic Feshbach resonances are available and
only moderate values of the scattering length can be reached~\cite{footnote_large_separations}.

%
%

\section{Modulation experiment} \label{sec:modulation_experiment}

Despite the intense investigation of the negative energy
bound states in recent spectroscopy experiments
with single layer systems~\cite{kohl_CIR, Sommer2012},
quasi-bound molecules remained elusive due to their very short lifetimes.
In bilayer gases considered here, however, interlayer quasi-bound molecules
can be made exponentially long-lived simply by increasing the layer separation,
and they can therefore be detected in simple shaking experiments.
To  demonstrate this, we determine the modulation spectrum of a strongly confined dilute
Bose gas (with a temperature $k_B T\ll \hbar \omega_z$),  
\marton{excited by the simultaneous shaking of both layers in opposite directions.
Such a shaking field can be conveniently produced either by applying a time dependent
magnetic field gradient, or through periodically modulated 
vector light shifts in a spin-dependent optical potential~\cite{spin_dependent_lattices1}.}
coupled to a time dependent magnetic field gradient that shakes 
the layers in opposite directions. 
To account for many-body effects, we describe the gas 
in terms of the second quantized Hamiltonian, 
\bea
H&=&\int {\rm d}^3 {\bf r} \bigl\{ \sum_\alpha\psi^\dagger_\alpha(\mathbf{r}) 
({\cal H}_{\alpha}-\mu_\alpha)
 \psi_\alpha (\mathbf{r}) 
 \label{eq:H_manybody} \\
&+&\, \sum_{\alpha,\beta} \frac{g_{\alpha\beta}}{2} \, \psi^\dagger_\alpha(\mathbf{r}) \psi^\dagger_\beta(\mathbf{r})
                                   \psi_\beta(\mathbf{r}) \psi_\alpha(\mathbf{r})
                                   \bigr\},
                                   \nonumber
\eea
with the fields $\psi_\alpha$ annihilating particles in layer $\alpha$, 
and the chemical potentials $\mu_\alpha<0$ setting the densities.
The interaction parameters $g_{\alpha\beta}$ are related to the three-dimensional
scattering lengths through appropriate renormalization~\cite{demler_pekker}.
Shaking is described by the modulation of the Hamiltonian
${\cal H}_{\alpha}$,  
\beq
\delta{\cal H}_\alpha(t) = - h_\alpha \cos{(\omega t)}\,  z_\alpha /l_z,
\label{eq:dH}
\eeq
with modulation frequency $\omega$, and the fields $h_{\alpha}$
characterizing the amplitudes of shaking for the two hyperfine components. 
Due to the selection rules imposed by harmonic confinement,
shaking induces $n \leftrightarrow (n+1)$ intralayer 
transitions within each layer, to leading order.
In a strongly confined Bose gas, dominantly $n=0 \rightarrow 1$ transitions will be excited,
since the $n>0$ levels are essentially unpopulated.
Decomposed in terms of center of mass ($N$) and relative ($\nu$) quantum numbers, 
these correspond to pair excitations $(N,\nu)=(0,0) \rightarrow (1,0)$ 
and $(0,0)\rightarrow(0,1)$~\cite{footnote5}.
Therefore, shaking not only allows to excite thermal particles to higher intrawell
bands (at energy $\hbar\omega_z$), but -- through the interaction with other thermal bosons -- 
it can also excite bound and quasi-bound molecular states.

\begin{figure}[t]
\includegraphics[width=8.6cm]{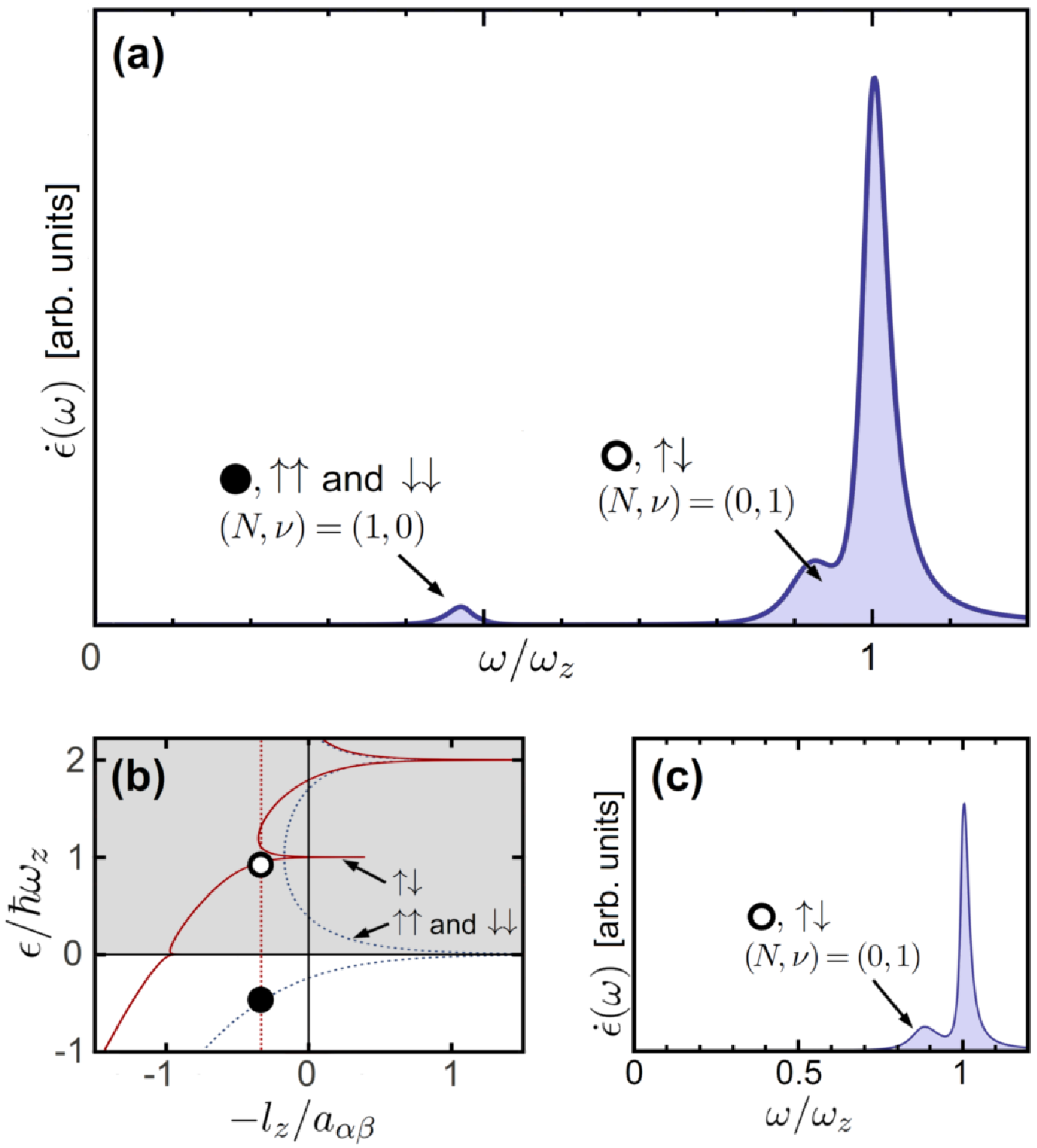}
\caption{(Color online.) 
(a) Shaking absorption spectrum at equal interaction strengths $a_{\alpha\beta}/l_z = 3$,
and layer separation $d_{\uparrow\downarrow}/l_z = 2.5$. 
Center of mass ($N$) and relative ($\nu$) quantum numbers
of the peaks related to bound and quasi-bound molecular states are indicated by
full and open circles, respectively, also shown in (b).
[Physical parameters: $|\mu_\uparrow| = |\mu_\downarrow| = k_BT /3, \, k_B T= 0.03\,\hbar \omega_z$.]
(c) In the absence of interlayer interactions ($a_{\upa\upa} = a_{\downa\downa} = 0$), 
the bound states in the $\upa\upa$ and $\downa\downa$-channels disappear,
and the corresponding peak vanishes from the modulation spectrum.
[Parameters of the inset: 
$a_{\uparrow\downarrow}=2.6 \,l_z,\; |\mu_\uparrow| = |\mu_\downarrow| = k_BT /3,\, k_BT=0.06 \,\hbar\omega_z$.]
} 
\label{fig:Spectrum}
\end{figure}

In particular, $(N, \nu)=(0,0) \rightarrow (0,1)$ transitions excite the $\uparrow\downarrow$ 
interlayer quasi-bound molecule of energy $E_{\upa\downa}^1$,
close to the $\hbar\omega_z$ threshold
(open circle in Fig.~\ref{fig:Spectrum}(b)).
The other, $(N,\nu)=(0,0)\rightarrow (1,0)$ excitation creates transitions 
to molecular bound states in the $\uparrow\uparrow$ and $\downa\downa$-channels.
Due to the center of mass energy $N\hbar \omega_z$, the energy of these
bound states can get shifted to positive values, $E_{\upa\upa}^B \rightarrow E_{\upa\upa}^B + \hbar \omega_z$ 
(and $E_{\downa\downa}^B + \hbar \omega_z$).
We therefore expect peaks at all these energies in the absorption spectrum~\cite{footnote4}.

\begin{figure}[b]
\includegraphics[width=8.6cm]{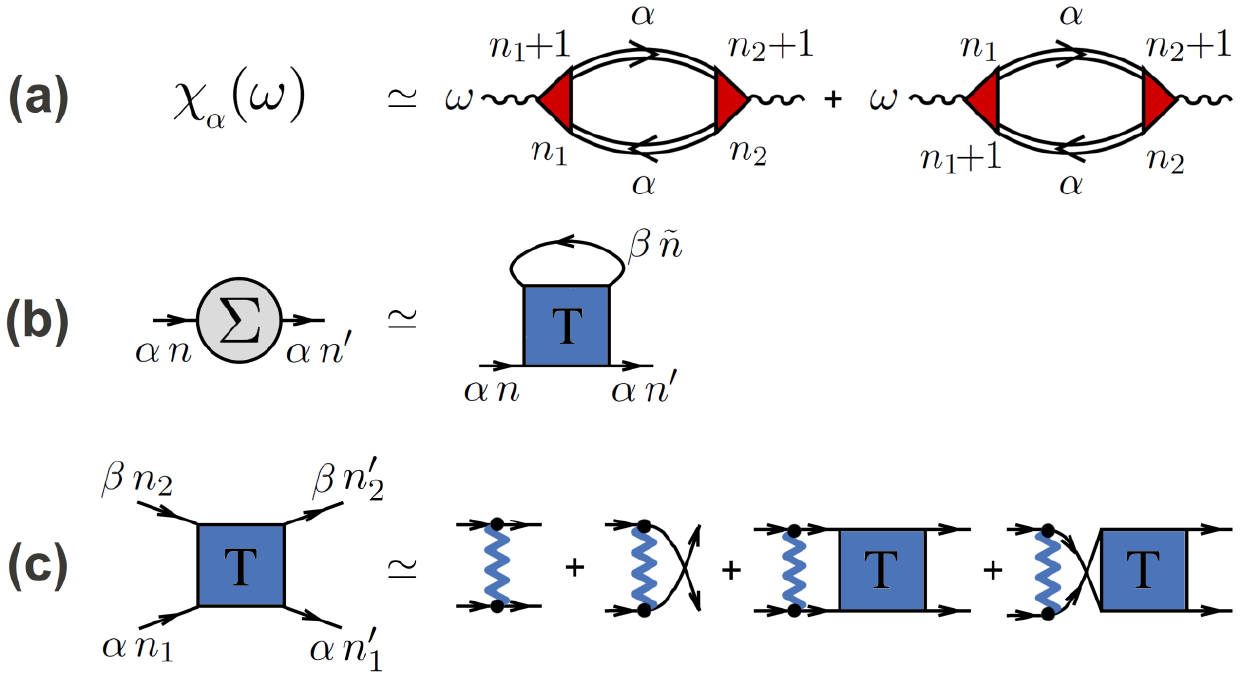}
\caption{(Color online.) 
{
Feynman diagrams determining the shaking spectrum:
(a) Shaking susceptibility, as calculated within linear response theory and neglecting vertex corrections.
Double lines indicate  propagators dressed by self-energy corrections (b), whereas  red triangles 
stand for  shaking vertices.
(b) Self-energy corrections within the $T$-matrix approximation.
(c) Bethe-Salpeter equations of the many-body $T$-matrix, approximated as a sum up ladder diagrams.
Full lines indicate bare propagators, dashed lines refer to the bare coupling.
}
}
\label{fig:FeynmanDiagrams}
\end{figure}

To verify these expectations, we calculated the imaginary part of 
the shaking susceptibility, $\chi^{\prime\prime}_\alpha (\omega)$, using field theoretical methods.
Neglecting vertex corrections, $\chi_\alpha(\omega)$ is given by the 'dressed' bubble diagrams
in Fig.~\ref{fig:FeynmanDiagrams}(a).
This quantity is directly related to the rate of energy absorption in layer $\alpha$, given by
$\dot{\epsilon}_\alpha(\omega) = h_\alpha^2 \, \omega \chi_\alpha^{\prime\prime}(\omega)/2$.
During their propagation, particles excited
by lattice modulations go through virtual transitions to bound and quasi-bound 
states with other particles in the thermal gas.
Interactions between the excited particles and the thermal gas are incorporated 
in the dressed propagators (heavy lines), through self-energy contributions.
To compute this, we expand the fields $\psi_\alpha$ in terms of harmonic oscillator wave functions.
$\psi_\alpha(\mathbf{r}) \propto  \sum_{n} \int d^2 q \;
{\varphi}_n(z-z^0_\alpha) \, e^{i\mathbf{q} \vec{\mathbf{\rho}}} \, a_{\alpha n}(\mathbf{q})$.
In this basis, the dressed retarded propagator of 
particles in layer $\alpha$ is given by
\beq
(G_R^{-1})^{n n'}_{\alpha} (\omega,\mathbf{q}) 
= \omega + i 0^+ + \frac{\hbar q^2}{2 m} + n \omega_z\, \delta_{n n'} + \frac{1}{\hbar}\Sigma^{n n'}_{\alpha} (\omega,\mathbf{q}),
\nonumber
\eeq
with the self-energy $\Sigma_{\alpha}^{n n'}$ accounting for interactions
with thermal particles (see Fig.~\ref{fig:FeynmanDiagrams}(b)), 
creating transitions between harmonic oscillator levels $n \rightarrow n'$.

We compute the self-energies 
within the $T$-matrix approximation by summing up the 
complete ladder diagrams for the $T$-matrix (vertex function), 
and solving the corresponding Bethe-Salpeter equations
(see Fig.~\ref{fig:FeynmanDiagrams}(c)). 
In the absence of thermal particles, the $T$-matrix approximation becomes exact, and
gives an expression identical to the scattering amplitudes in Eq.~\eqref{eq:T-matrix_2body}, up to a normalizing constant.
In a dilute Bose gas, however, the $T$-matrix contains additional many-body contributions, 
accounting for screening effects~\cite{demler_pekker}.
Details of these many-body calculations are given in 
Appendix~\ref{app:shaking}, here we just summarize the main results.

Fig.~\ref{fig:Spectrum}(a) displays the numerically computed shaking spectrum for some typical 
parameters in the cross-over regime, $a_{\alpha\beta}\sim l_z$ and $d_{\uparrow\downarrow}\sim l_z$. 
For simplicity, we assumed shaking fields $h_\uparrow = h_\downa $, and  repulsive three-dimensional 
scattering lengths of equal size in all three scattering channels, $a_{\alpha\beta}=a$.  
Three peaks are clearly distinguishable in the spectrum. The largest peak at $\omega \approx \omega_z$ corresponds
the direct subband transitions within the same layer, and has an amplitude directly proportional to the 
boson density. We also observe, however, two smaller peaks. 
These are due to two-body processes, therefore their intensities are proportional 
to the \emph{square} of the boson densities. 
The peak next to the large quasiparticle excitation peak is due to the quasi-bound molecular state 
in the $\upa\downa$-channel at energy $E_{\upa\downa}^1$, indicated by the empty circle in Fig.~\ref{fig:Spectrum}(b).
As expected, for separations 
$d\sim l_z$  this peak is indeed sharp enough and  can be  identified unambiguously within the 
shaking spectrum. Although this quasi-bound resonance may appear  relatively  weak at a first sight 
for the thermal gas studied here, it is expected to get more pronounced at higher densities, as 
the system is driven towards quantum degeneracy -- a regime beyond the reach of the approximations used here. 

The  peak at even smaller frequencies has an entirely 
different origin, and is attributed to
a  transition into the $\upa\upa$ or $\downa\downa$ 
intralayer bound states combined with a center of mass excitation, $N=0\to 1$ (full circle in Fig.~\ref{fig:Spectrum}(b)). 
Notice that -- to leading order -- a direct transition to the bound state due to shaking 
is forbidden by symmetry (parity), and therefore excitation of a center of mass
oscillation is necessary to observe the $E^0_{\upa\upa}$ and 
$E^0_{\downa\downa}$ bound states.
This peak is expected to split up for $a_{\uparrow\upa}\ne a_{\downa\downa}$, 
and it vanishes if we set $a_{\uparrow\upa}, a_{\downa\downa}\to 0$
(see Fig.~\ref{fig:Spectrum}(c)). 
The bound state in the $\upa\downa$ scattering gives a tiny contribution for these parameters, 
and is practically not visible in Figs.~\ref{fig:Spectrum}(a, c).  

\marton{
Finally, we mention that time-modulation of the scattering lengths instead of the layer separation
may provide an alternative experimental procedure of exciting the bound and quasi-bound molecular
states, most probably leading to smaller quasi-particle contributions.
Unlike the modulation of the layer separation, this perturbation couples the 
$(N,\nu) = (0,0)$ state to a large number of harmonic oscillator channels, and it 
is therefore likely to excite a number of molecular states at higher energies as well.
}
%
%

\section{Conclusions and outlook on experiments} \label{sec:conclusion}

In the preceding sections, we have studied theoretically  how confinement modifies interactions
in bilayer gases of ultracold atoms. 
We determined the two-particle scattering amplitudes and 
demonstrated the existence of  confinement-induced interlayer molecular bound states  for all values of 
the the scattering lengths $a_{\upa\downa}$. 
Rather counter-intuitively, these exciton-like interlayer molecular states exist even at layer separations
several times larger than the oscillator length $l_z$ of the confining potential, 
in a regime, where the overlap between the clouds is almost negligible.

At positive energies, $\epsilon>0$, the scattering amplitudes exhibit confinement induced 
resonances of finite width, attributed to quasi-bound molecular states of finite lifetime.
Here we have also demonstrated   
the existence of  a   novel kind of quasi-bound interspecies molecular states
at energies  $\epsilon \simeq \hbar\omega_z$, 
absent in single layer systems. 
These resonances are due to virtual processes,
whereby two colliding atoms bind into a virtual interlayer molecular state, which then decays into
the continuum.
The energy and lifetime of these quasi-bound molecular states depend sensitively 
on the layer separation $d_{\upa\downa}$ and on the 
scattering lengths $a_{\upa\downa}$  (see Sec.~\ref{sec:interaction_control}). 
The sensitivity of these  novel  resonances can be exploited to \emph{engineer} the interaction between the 
two species: rather counterintuitively, 
the interaction between two species of atoms can be  made  resonant by spatially  \emph{separating}  the two species.  
The geometrical interaction control   proposed here is efficient also for 
moderate values of the scattering lengths $a_{\upa\downa}$, can therefore be used in
a range of atomic species,
and thus  paves the way to realizing   novel strongly-correlated many-body phases.

Ordinary confinement induced resonances in quasi-two-dimensional systems 
are extremely broad and, in fact, have never been detected before. 
The novel interspecies resonance, however,  
is getting  sharper as the two species are separated and, as we have 
demonstrated for the case of a strongly confined 
dilute Bose gas in Sec.~\ref{sec:modulation_experiment}, 
it is   observable in a simple shaking experiment, where it appears as a clearly 
distinguishable absorption peak. 
Due to its two-body character, the intensity of the corresponding absorption  peak
is proportional to the square of the boson density, and 
is expected to become more pronounced as the system is driven
towards quantum degeneracy.

%

%
%
For an experimental observation  of the above effects, 
one needs to reach the regime, where all natural length scales are of the same order, 
$a_{\alpha\beta} \sim  l_z \sim d_{\upa\downa} $. 
One way to reach this regime is to decrease $l_z$, by applying 
extremely strong trapping frequencies, or by using 
heavy atoms (such as $\mathrm{^{87} Rb}$)~\cite{footnote6}.
No such a strong confinement is needed, however, for  atoms with interspecies Feshbach 
resonances, where the scattering length $a_{\upa\downa}$ can 
be tuned by magnetic fields to large enough values  at standard trapping 
frequencies,  $\omega_z = 10-100~\mathrm{kHz}$~\cite{footnote__experimental_parameters}.
%
%
Although in case of bosonic species, broader resonances between hyperfine 
levels of the same atom are relatively rare~\cite{chin_review, Rb_interspecies_resonance},
they are much more common and widely used
in the case of bosonic mixtures, such as $^7\mathrm{Li}$--$^{87}\mathrm{Rb}$, 
$^{39}\mathrm{K}$--$^{87}\mathrm{Rb}$ and $^{41}\mathrm{K}$--$^{87}\mathrm{Rb}$
systems~\cite{chin_review,bosonic_mixtures_1,bosonic_mixtures_2,bosonic_mixtures_3,bosonic_mixtures_4}.
Although for species of unequal masses the center-of mass and relative motion decouple only for 
equal confinement frequencies, we do not expect our results to change dramatically even if these two confinement frequencies are 
different~\cite{footnote7}. In these systems the regime $a_{\upa\downa}\sim l_z$ is thus readily accessible.

%
%
The ideas presented here are not limited to bosonic systems. 
Our results on two-particle scattering in the vacuum, discussed in Sec.~\ref{sec:two-particle}, 
apply to dilute Fermi gases as well. In the Fermionic case,
s-wave scattering between identical fermion species is inactive due to the Pauli principle ($a_{\upa\upa}, a_{\downa\downa}\to 0$).
Since fermionic systems (such as $^6\mathrm{Li}$ and $^{40}\mathrm{K}$) 
have sufficiently broad and widely used Feshbach resonances, the regimes required to observe the 
effects of interlayer quasi-bound states can be easily reached within standard experiments.
Fermionic gases are therefore promising candidates for detecting interlayer quasi-bound molecules
in modulation experiments,  and for implementing the  geometrical interaction control discussed here.
In the fermionic case, many-body effects can be accounted for by similar methods to those presented in 
Sec.~\ref{sec:modulation_experiment}~\cite{demler_pekker}, and
could  lead to several exotic phenomena in a Fermi degenerate gases
such as exciton condensation~\cite{Butov_Nature2002,SeamonsLilly_PRL2009}.

\section{Acknowledgements}
We would like to thank M. Zwierlein, J. Dalibard, M. Greiner, W. Ketterle, and  M. Babadi   for illuminating discussions. 
This research has been supported  by the Hungarian Research Funds under grant Nos. K105149, CNK80991.
E.A.D. Acknowledges support through  the Harvard-MIT CUA, the DARPA OLE
program,  the AFOSR MURI on Ultracold Molecules,  and  the ARO-MURI on Atomtronics projects.
\null

\appendix

\section{Short distance asymptotics of the retarded Green's function
\label{app:Greens_function}}

We determine the constant part of  $G^{(0)}_\epsilon$
at short distances, by comparing Eqs.~\eqref{eq:retarded_propagator} and \eqref{eq:G_short_distance},
leading to
\beq
w_{\alpha\beta} =
\lim_{\rho \to 0} \left( 2 \sum_{\nu=0}^\infty \frac{\phi_{\nu}^2(d_{\alpha\beta})}{\phi_0^2(0)}
K_0\left( -i q_\nu \rho \right)
- \frac{\sqrt{2\pi}l_z}{\rho}\right),\label{eq:w_rho}
\nonumber
\eeq
with $\phi_0^2(0) = 1 / (\sqrt{2\pi} l_z)$.
To simplify this expression, we choose a large 
integer, $\overline{\nu} \gg 1$, and split the sum above into two parts, 
with $\nu < 2\overline{\nu}$ and $\nu \ge 2 \overline{\nu}$.
We assume that $\rho$ is already small, so that $\kappa \equiv \sqrt{2 \overline{\nu}} \rho/l_z$ is a small parameter.
In the $\nu< 2\overline{\nu}$ part of the sum, 
the Bessel function can be approximated by its asymptotic form
\beq
K_0(x\rightarrow 0)\sim -\log(x/2) - \gamma_E.
\label{eq:K_0_asymptotic}
\eeq
In the $\nu \ge 2 \overline{\nu}$ part, on the other hand,
$K_0$'s argument is well approximated by $\sqrt{\nu}\rho/l_z$, and we can
make use of the asymptotic form of the Hermite 
functions in the limit $\nu\to\infty$,
\beq
\frac{\phi_{\nu}^2(d_{\alpha\beta})}{\phi_0^2(0)}
\sim \sqrt{\frac{2}{\pi}} \, 
\frac{\cos^2{\left(\frac{d_{\alpha\beta}}{l_z} \sqrt{\nu+\frac{1}{2}} - \nu \frac{\pi}{2}\right)}}{\sqrt{\nu}}.
\nonumber
\eeq
As $\nu$ varies in this part of the sum, the $\cos^2$ term averages 
out to $1/2$, whereas $K_0$'s argument changes only slowly, 
$\rho/l_z$ being a small parameter.
Thus, the $\nu\ge 2\overline{\nu}$ part of the sum in Eq.~\eqref{eq:w_rho} 
can be approximated by an integral,
\beq
\begin{aligned}
&\sum_{\nu=2\overline{\nu}}^\infty \frac{\phi_{\nu}^2(d_{\alpha\beta})}{\phi_0^2(0)}
K_0\left( -i q_\nu \rho \right) \nonumber \\
&\simeq
\sum_{\nu=2\overline{\nu}}^\infty \frac{1}{\sqrt{2\pi \nu}} \, K_0(\sqrt{\nu} \rho / l_z) \\
&\simeq
\frac{l_z}{\rho} \sqrt{\frac{2}{\pi}} \int_{\kappa}^\infty dx \, K_0(x)
\\
&\simeq
\frac{l_z}{\rho} \sqrt{\frac{2}{\pi}} 
\left( \frac{\pi}{2} + \kappa \left(\log{\frac{\kappa}{2}} + \gamma_E - 1 \right) \right),
\end{aligned}
\nonumber
\eeq
with $x=\sqrt{\nu}\,\rho/l_z$,
where we made use of the formula $\int_0^\infty K_0(x) \, dx = \pi/2$ and of
the asymptotic form of $K_0$ in Eq.~\eqref{eq:K_0_asymptotic}.
Finally, by putting the two parts of the sum together, we can take the limits 
$\rho\to 0$ and $\nu\to\infty$ (by keeping $\kappa\rightarrow 0$), and get
\beq
\begin{aligned}
w_{\alpha\beta} &= \lim_{\overline{\nu}\to\infty} \left[ c_{\overline{\nu}}
- \sum_{\nu=0}^{2\overline{\nu} - 1} \frac{\phi_{\nu}^2(d_{\alpha\beta})}{\phi_0^2(0)} 
\log{\left(\frac{\nu}{2}-\frac{\epsilon + i 0^+}{\hbar\omega_z}\right)}
\right.
\\
&+\left.
\left( \log{(\rho/\sqrt{2}l_z)} + \gamma_E \right)
\left( 4\sqrt{\frac{\overline{\nu}}{\pi}} 
- 2 \sum_{\nu=0}^{2\overline{\nu} -1} \frac{\phi_{\nu}^2(d_{\alpha\beta})}{\phi_0^2(0)}  \right)
\right],
\end{aligned}
\nonumber
\eeq
with $c_\nu=2 \sqrt{\frac{\nu}{\pi}} \log{\frac{\nu}{e^2}}$.
In the $\overline{\nu}\to \infty$ limit the term in the second row above 
disappears, and we get back the desired form of $w_{\alpha\beta}$, 
as given below Eq.~\eqref{eq:T-matrix_2body} in the main text.

%
%
%
%
This series representation of $w_{\alpha\beta}$, however,
has particularly poor $\sim \log\overline{\nu}/\sqrt{\overline{\nu}}$ convergence properties, 
and also oscillatory behavior in the $d_{\upa\downa}\ne 0$ case,
that make it impractical for numerical evaluations.
In the following, we thus provide an integral representation of this expression,
which is more useful for numerical applications.
Generalizing the calculations of Ref.~\onlinecite{demler_pekker}, 
we first rewrite the terms in Eq.~\eqref{eq:w_rho} in an integral representation,
\beq
\frac{\sqrt{2\pi} l_z}{\rho} = \frac{1}{\sqrt{2}}
\int_0^\infty \frac{d\tau}{\tau^{3/2}} e^{-\rho^2/(4 l_z^2\, \tau)},
\nonumber
\eeq
and
\beq
\begin{aligned}
\frac{K_0\left( -i q_{\nu}\rho \right)}{2\pi} &= -\frac{\hbar^2}{m}\int \frac{d^2 k}{(2 \pi)^2} 
\frac{e^{i\mathbf{k \rho}}}{\epsilon + i 0^+ - \left( \frac{\hbar^2 q^2}{m} + \hbar \nu \omega_z \right)} 
\\
&= \int_0^\infty \frac{d\tau}{4 \pi \tau} \, e^{\tau \left(\epsilon/\hbar\omega_z - \nu\right)} e^{-\frac{\rho^2}{4 \, l_z^2 \, \tau}},
\end{aligned}
\nonumber
\eeq
that holds for all values of $\nu$ above the threshold $\nu > \epsilon/\hbar\omega_z$.
Let us thus choose an arbitrary integer $\hat{\nu} > \epsilon / \hbar \omega_z$, 
and rewrite the terms $\nu > \hat{\nu}$ in Eq.~\eqref{eq:w_rho} in the above form,
leading to
\beq
\begin{aligned}
w_{\alpha\beta}
&= \lim_{\rho \to 0} \left\{ 2 \sum_{\nu=0}^{\hat{\nu}}
\frac{\phi_{\nu}^2(d_{\alpha\beta})}{\phi_0^2(0)}
K_0\left(-i q_{\nu}\rho\right) \right.
\\
&+\int_0^\infty \frac{d\tau}{\tau} e^{-\rho^2/(4 l_z^2\, \tau)}
\left[
-\frac{1}{\sqrt{2 \tau}}
\right.
\\
&\left.\left.
+\sum_{\nu = \hat{\nu}+1}^\infty 
\frac{\phi_{\nu}^2(d_{\alpha\beta})}{\phi_0^2(0)}
e^{\tau (\epsilon / \hbar\omega_z -\nu)}
\right]\right\}.
\end{aligned}
\label{eq:w_aux}
\eeq
The infinite sum above can be carried out exactly by making use of the
formula for the real space density matrix of a harmonic 
oscillator~\cite{feynman_statistical_mechanics},
\beq
\sum_{\nu=0}^\infty \frac{\phi_{\nu}^2(\mathrm{z})}{\phi_0^2(0)} \, e^{-\tau \nu}
= \sqrt{\frac{e^{\tau}}{2 \sinh{ \tau}}}e^{-\tanh{\left(\tau/2\right)}\, \mathrm{z}^2/2 l_z^2}.
\nonumber
\eeq
In order to take the $\rho \to 0$ limit, we expand the Bessel function, $K_0$, up to
linear order in $\rho$, and rewrite its $\log ( \rho )$ singularity in an integral form,
\beq
\begin{aligned}
K_0\left(-i q_{\nu}\rho\right) &\sim
\left[ i\pi - \gamma_E - \log{\left( \frac{\nu\hbar\omega_z - \epsilon - i 0^+}{4\hbar\omega_z}\right)} \right.
\\
&+ \left. \int_{0}^\infty \frac{d\tau}{\tau} \, \Theta\left(\frac{1}{4}-\tau\right) \, e^{-\rho^2/(4 l_z^2\, \tau)}\right]/2,
\end{aligned}
\nonumber
\eeq
with the Heaviside function $\Theta(\tau)$ and Euler's constant $\gamma_E\approx 0.577$.
Substituting these expressions into Eq.~\eqref{eq:w_aux}, we can take the
$\rho\to 0$ limit, and write $w_{\alpha\beta}$ in the form
\beq
\begin{aligned}
w_{\alpha\beta} &=
-\sum_{\nu=0}^{\hat{\nu}}
\frac{|\phi_{\nu}(d_{\alpha\beta})|^2}{|\phi_{0}(0)|^2}
\left[\log{\left( \frac{ \nu\hbar\omega_z -\epsilon - i 0^+}{4\hbar\omega_z}\right) } + \gamma_E \right]
\\
&+\int_0^\infty\frac{d \tau}{\tau} 
\left[ 
e^{-\frac{\tau \epsilon}{\hbar\omega_z}} 
\sqrt{\frac{e^\tau}{2 \sinh{\tau}}} e^{-\tanh{\left(\tau/2\right)}\, d_{\alpha\beta}^2/2 l_z^2} - \frac{1}{\sqrt{2\tau}} 
\right.
\\
&+
\left.
\sum_{\nu=0}^{\hat{\nu}} \frac{|\phi_{\nu}(d_{\alpha\beta})|^2}{|\phi_{0}(0)|^2} 
\left( \Theta\left(\frac{1}{4}-\tau\right) - e^{\tau\left(\epsilon/\hbar\omega_z - \nu\right)} \right)
\right].
\end{aligned}
\nonumber
\eeq
Despite its complexity at first glance, this formula provides a simple and fast numerical method 
for calculating $w_{\alpha\beta}$, and we have used it to evaluate the scattering amplitudes, $f_{\alpha\beta}^{\nu\nu'}$,
to high numerical accuracy.

\section{Shaking experiment} \label{app:shaking}

In order to separate the motional degrees of freedom parallel and perpendicular to the 
two-dimensional planes, we rewrite the many-body Hamiltonian in Eq.~\eqref{eq:H_manybody}
in terms of the annihilation operators
\beq
a_{\alpha n}(\mathbf{q}) = \int d^2 \rho \int d z \, e^{-i \mathbf{q} \vec{\mathbf{\rho}}} \,
\varphi(z-z_\alpha^0) \, \psi_\alpha(\mathbf{r}).
\nonumber
\eeq
The normalization of these operators is given by their commutation relations
$[ a_{\alpha n}(\mathbf{q}),a^\dagger_{\alpha' n'}(\mathbf{q'}) ] 
= (2\pi)^2 \delta_{\alpha\alpha'} \, \delta_{n n'} \, \delta^{(2)} (\mathbf{q} - \mathbf{q}')$.
In this basis, the many-body Hamiltonian $H = H_{\mathrm{kin}} + H_{\mathrm{int}}$ can be written
in the form
\beq
\begin{aligned}
H_{\mathrm{kin}} =& \sum_{\alpha=\uparrow,\downarrow} \sum_{n=0}^\infty \int\frac{d^2 q}{(2\pi)^2} \,
\xi_{\alpha n} (\mathbf{q}) \, a^\dagger_{\alpha n} (\mathbf{q}) a_{\alpha n} (\mathbf{q}),
\\
H_{\mathrm{int}} =& \sum_{\alpha, \beta=\uparrow,\downarrow} \sum_{\mathbf{n},\mathbf{n'}} 
\int \frac{d^2 k}{(2\pi)^2} \frac{d^2 k'}{(2\pi)^2} \frac{d^2 q}{(2\pi)^2} \;
\frac{t_{\alpha\beta}^{\mathbf{n} \mathbf{n'}}}{2}
\\
& a^\dagger_{\alpha n_1} (\mathbf{k+q}) a^\dagger_{\beta n_2} (\mathbf{k'-q})
a_{\beta n_2'} (\mathbf{k'}) a_{\alpha n_1'}(\mathbf{k}),
\end{aligned}
\nonumber
\eeq
where $\xi_{\alpha n}(\mathbf{q})=\hbar^2 q^2 / 2 m + n \hbar \omega_z - \mu_\alpha $
stands for the single particle energies measured from the corresponding chemical potential $\mu_\alpha$,
and $\mathbf{n}=(n_1,n_2)$ denotes the harmonic channels of the interacting particles.
The interaction parameter $\mathbf{t}_{\alpha\beta}^{\mathbf{n} \mathbf{n'}}$
is the bare $T$-matrix (vertex) of interactions, and it is given by
\beq
\begin{aligned}
\mathbf{t}_{\alpha\beta}^{\mathbf{n} \mathbf{n'}} &= 
\,g_{\alpha\beta} \int_{-\infty}^\infty d z_1 \, d z_2 \, \delta(z_1-z_2 + d_{\alpha\beta})\\
&\hspace{50 pt} \varphi_{n_1}^*(z_1) \varphi_{n_2}^*(z_2) \varphi_{n_2'}(z_2) \varphi_{n_1'}(z_1) \\
&= g_{\alpha\beta} \sum_{N \nu \nu'} C^{\mathbf{n} \, *}_{N \nu} \, C^{\mathbf{n}' }_{N \nu'} \, 
\phi^*_\nu(d_{\alpha\beta})\phi_{\nu'}(d_{\alpha\beta}),
\end{aligned}
\nonumber
\eeq
where the Clebsch-Gordan coefficients $C^{\mathbf{n}}_{N \nu}$ denote the overlaps 
$C^{\mathbf{n}}_{N \nu} = \langle N \nu | \mathbf{n} \rangle$.

Assuming an equal coupling of the magnetic field gradient to the 
spin components $h_\upa = h_\downa = h_0$ in Eq.~\eqref{eq:dH}, 
shaking of the layers is described by the modulation Hamiltonian 
$\delta H_\alpha = - h_0 \cos(\omega t) \, \Xi_\alpha$, 
with $\Xi_\alpha = z_\alpha / l_z$ given in many-body form as
\beq
\Xi_\alpha = \sum_{n=0}^\infty \int \frac{d^2 q}{(2\pi)^2} 
\sqrt{\frac{n+1}{2}} \left( a_{\alpha n+1}^\dagger (\mathbf{q}) \, a_{\alpha n} (\mathbf{q})
+ \mathrm{h.c.} \right).
\nonumber
\eeq
Thus, the shaking excites $n \leftrightarrow (n+1)$ transitions in both layers, 
amounting to $n=0 \to 1$ transitions in case of a strongly confined dilute Bose gas.
In linear response theory, the shaking susceptibility is given by the Kubo formula
\beq
\chi_\alpha(t) = i \Theta (t) \, \langle \left[\Xi_\alpha(t),\Xi_\alpha(0)\right]\rangle,
\nonumber
\eeq
which is approximated by taking into account the bubble diagrams in
Fig.~\ref{fig:FeynmanDiagrams}(a) with dressed propagators, as we explained in the main text.
The self-energy corrections to the propagator are due to the interaction of the 
propagating particles with the thermal gas through the many-body $T$-matrix.

In the the diagrammatic approach, incoming particles are specified by their frequency $\omega$, 
momentum ${\bf q}$, layer index $\alpha$, and their transverse channel $n$.  
The two particle $T$-matrix corresponds to the  vertex function within the field theoretical approach, 
and in the vacuum it is proportional to the scattering amplitudes in Eq.~\eqref{eq:T-matrix_2body}
with the total energy of the incoming bosons replaced by the 
sum of their frequencies,  $\Omega = \omega_1 + \omega_2$ (see Ref.~\onlinecite{demler_pekker}).
Both $\Omega$ and the total incoming momentum $\mathbf{Q} = \mathbf{q}_1 + \mathbf{q}_2$ 
are conserved within the 'ladder' diagram approximation of Fig.~\ref{fig:FeynmanDiagrams}(c), 
unlike $n_1$ and $n_2$, which are not conserved. However, 
similar to the two particle problem~\cite{demler_pekker},
it is possible to sum up the whole ladder series by transforming to center of 
mass and relative coordinates and to the corresponding  quantum numbers, 
$\{n_1,n_2\}\to \{N,\nu\} $. The 
total  many-body  vertex in Fig.~\ref{fig:FeynmanDiagrams}(c) can then be expressed as 
\bea
\mathbf{T}_{\alpha\beta}^{\mathbf{n} ; \mathbf{n}^\prime}(\Omega,\mathbf{Q}) 
&=& \sum_{N,N',\nu,\nu'}
C_{N \nu}^{\mathbf{n} \; *}\; C_{N' \nu'}^{\mathbf{n}^\prime}\;
\mathbf{T}_{\alpha\beta}^{NN'; \nu,\nu^\prime }(\Omega,\mathbf{Q}),
\nonumber
\eea
with the Clebsch-Gordan coefficients $C_{N \nu}^{\mathbf{n}}$ defined in Appendix~\ref{app:shaking}.
In the strongly confined Bose gas, where only the lowest $n=0$
level is populated, the $T$-matrix becomes diagonal in the center of mass index, 
$\mathbf{T}_{\alpha\beta}^{NN'; \nu,\nu^\prime }\to  \delta_{N N'}\mathbf{T}_{\alpha\beta}^{N; \nu,\nu^\prime }$,
and is given by
\beq
\mathbf{T}_{\alpha\beta}^{N; \nu,\nu^\prime }(\Omega,\mathbf{Q}) = 
\frac{\hbar^2}{m} \;
\frac{
  4 \pi a_{\alpha\beta} \, 
  \phi^*_{\nu}(d_{\alpha\beta}) \phi_{\nu'}(d_{\alpha\beta})
  }{
  1 + \frac{a_{\alpha\beta}}{\sqrt{2 \pi} l_z} 
  \mathcal{W}^N_{\alpha\beta}(\Omega,\mathbf{Q})
}, \nonumber
\eeq
with the many-body counterpart $\mathcal{W}_{\alpha\beta}^N$ of
$w_{\alpha\beta}$,
\beq
\mathcal{W}^N_{\alpha\beta}(\Omega,\mathbf{Q}) = w_{\alpha\beta}\left( 
\epsilon/\hbar\omega_z 
\right)
+ \delta_{N 0} \; \delta w_{\alpha\beta}^\mathrm{th}(\Omega,\mathbf{Q}), \nonumber
\eeq
with $\epsilon=\hbar \Omega -N \hbar \omega_z - \hbar^2 Q^2/4 m $.
The first term in $\mathcal W$ is just the vacuum 
contribution computed earlier, while the second term accounts for many-body interactions 
with other thermal bosons, and it is proportional to the density,
\beq
\delta w^\mathrm{th}_{\alpha\beta}(\Omega,\mathbf{Q})=
- \frac{4 \pi}{m} \sqrt{2\pi} l_z \, |\phi_{0}(d_{\alpha\beta})|^2 \, 
\Pi_{\alpha\beta}^\mathrm{th}(\Omega,\mathbf{Q}),
\nonumber
\eeq
with
\beq
\Pi^{\mathrm{th}}_{\alpha\beta} (\Omega,\mathbf{Q})  =
\sum_{\gamma=\alpha,\beta}\int \frac{d^2 q}{(2\pi)^2} 
\frac{
  n_B\left(\frac{\hbar^2(\frac{\mathbf{Q}}{2} + \mathbf{q})^2}{2m}-\mu_\gamma\right)
  }{
  \frac{\Omega + i 0^+}{\hbar} - \frac{Q^2}{4m} - \frac{q^2}{m}
  },\nonumber
\eeq
where $n_B$ denotes the Bose distribution function.
$ \delta w^\mathrm{th}_{\alpha\beta}$ describes the screening effect of the thermal gas.
In case of a dilute Bose gas, however, its contribution to the dressed Green's 
functions turns out to be numerically small,
and most features observed in the shaking experiment are dominated 
by the vacuum scattering amplitudes, determined by just $w_{\alpha\beta}$.  
Neglecting thermal corrections, the $T$-matrix becomes proportional 
to the vacuum scattering amplitudes 
\beq
T_{\alpha\beta}^{N;\nu\nu'}(\Omega,\mathbf{Q}) \approx \frac{\hbar^2}{m} 
f_{\alpha\beta}^{\nu\nu'} \left( \hbar\Omega - N\hbar\omega_z - \frac{\hbar^2 Q^2}{4 m} \right).
\nonumber
\eeq

In calculating the self-energy, $\Sigma_\alpha^{n n'}$, due to the diluteness of the gas,
we only keep terms proportional to the square
of the density, leading to
\beq
\begin{aligned}
\Sigma^{n n'}_{ \alpha}(\omega,{\bf q}) \approx&\sum_{\beta=\uparrow,\downarrow}\sum_{\tilde n=0}^\infty \int \frac{d^2 k}{(2\pi)^2} \,
n_B\left(\frac{\hbar^2\mathbf{k}^2}{2 m} + \tilde{n} \hbar \omega_z - \mu_\beta\right)\\
&\mathbf{T}_{\alpha\beta}^{n \tilde n,n' \tilde n}\left(\omega + \frac{\hbar \, k^2}{2 m} 
 +\tilde n\omega_z,\mathbf{k}+\mathbf{q}\right).
 \nonumber
\end{aligned}
\eeq
Having the self-energies at hand, we can proceed and compute the spectral functions of the Green's functions, 
$\rho^{n n'}_\alpha(\omega,\mathbf{q})=- \mathrm{Im}(G_R)^{\,n n'}_{\alpha} (\omega,\mathbf{q})/\pi$.
In terms of the spectral functions, the shaking susceptibility takes on a particularly simple form.
In the strongly confined gas only the lowest $n=0,1$ levels give dominant contributions to the susceptibility, thus we obtain
\beq
\begin{aligned}
\chi_\alpha(\omega)=&\int \frac{d\tilde{\omega}}{2}\frac{d^2 \tilde{q}}{(2 \pi)^2} \, n_B(\tilde{\omega}-\mu_\alpha)
\,\Bigl[ \rho^{00}_{\alpha}(\tilde{\omega},\tilde{q}) \rho^{11}_{\alpha}(\tilde{\omega}+\omega,\tilde{q}) \\
&+ 2\,\rho^{01}_{\alpha}(\tilde{\omega},\tilde{q})\, \rho^{01}_{\alpha}(\tilde{\omega}+\omega,\tilde{q})
+\{\omega \leftrightarrow -\omega\}\Bigr]\;.
\nonumber
\end{aligned}
\eeq
We determine the above integrals numerically, and arrive at the shaking spectrum 
$\dot{\epsilon}(\omega) = \sum_\alpha h_\alpha^2 \, \omega \chi_\alpha^{\prime\prime}(\omega)/2$,
shown in Fig.~\ref{fig:Spectrum}.


\begin{thebibliography}{}
\bibitem{fractional_q_hall} C.~Nayak, S.~H.~Simon, A.~Stern, M.~Freedman, and S.~Das~Sarma, Rev. Mod. Phys. {\bf 80}, 1083 (2008).
\bibitem{q_spin_hall_1} F.~D.~M.~Haldane, Phys. Rev. Lett. {\bf 61}, 2015 (1988).
\bibitem{q_spin_hall_2} B.~A.~Bernevig, T.~L.~Hughes, and S.-C.~Zhang, Science {\bf 314}, 1757 (2006).
\bibitem{high_Tc_review} P.~A.~Lee, N.~Nagaosa, and X.-G.~Wen, Rev.~Mod.~Phys. {\bf 78}, 17 (2006).
\bibitem{dalibard_BKT} Z.~Hadzibabic, P.~Kr\"uger, M.~Cheneau, B.~Battelier and J.~Dalibard,
Nature {\bf 441}, 1118 (2006).
\bibitem{nagerl_TG_gas} E.~Haller {\it et al.}, Science {\bf 325}, 1224 (2009).
\bibitem{ketterle_pencake_trap} A.~G\"orlitz {\it et al.}, Phys. Rev. Lett. {\bf 87}, 130402 (2001).
\bibitem{nagerl_cigar_shaped} E.~Haller {\it et al.}, Phys. Rev. Lett. {\bf 104}, 153203 (2010).
\bibitem{hermite_gauss} N.~L.~Smith, W.~H.~Heathcote, G.~Hechenblaikner, E.~Nugent, and C.~J.~Foot,
J.~Phys.~B: At.~Mol.~Opt.~Phys. {\bf 38}, 223 (2005).
\bibitem{chin_BKT} C.-L.~Hung, X.~Zhang, N.~Gemelke, and C.~Chin, Nature {\bf 470}, 236 (2011).
\bibitem{dalibard_BKT_PRL} P.~Kr\"uger, Z.~Hadzibabic, and J.~Dalibard, Phys. Rev. Lett. {\bf 99}, 040402 (2007).
\bibitem{phillips_BKT} P.~Clad\'e, C.~Ryu, A.~Ramanathan, K.~Helmerson, and W.~D.~Phillips, Phys. Rev. Lett. {\bf 102}, 170401 (2009).
\bibitem{krauth_BKT} M.~Holzmann and W.~Krauth, Phys. Rev. Lett. {\bf 100}, 190402 (2008).
\bibitem{feynman_q_simulators} R.~P.~Feynman, Int.~J.~Theor.~Phys. {\bf 21}, 467–488 (1982).
\bibitem{zwierlein__feynman_emulator} K.~Van~Houcke {\it et al.}, Nat. Phys. {\bf 8}, 366 (2012).
\bibitem{Butov_Nature2002} L. V. Butov, A. C. Gossard, and D. S. Chemla, 
Nature {\bf 418}, 751 (2002).
\bibitem{SeamonsLilly_PRL2009} J.~A.~Seamons, C.~P.~Morath, J.~L.~Reno, and M.~P.~Lilly, 
Phys. Rev. Lett. {\bf 102}, 026804 (2009). 
\bibitem{TutucHuse_PRL2004} E. Tutuc, M. Shayegan, and D. A. Huse,
Phys. Rev. Lett. {\bf 93}, 036802 (2004).
\bibitem{KeloggEisensteinWest_2004} M. Kellogg, J. P. Eisenstein, L. N. Pfeiffer, and K. W. West, 
Phys. Rev. Lett. {\bf 93}, 036801 (2004).
\bibitem{EisensteinMacdonald2004}
J. P. Eisenstein and A. H. MacDonald, Nature {\bf 432}, 691 (2004).
\bibitem{SuenTsui1992} Y. W. Suen, L. W. Engel, M. B. Santos, M. Shayegan, and D. C. Tsui,
Phys. Rev. Lett. {\bf 68}, 1379 (1992). 
\bibitem{EisensteinHe1992}
J. P. Eisenstein, G. S. Boebinger, L. N. Pfeiffer, K. W. West, and S. He,
Phys. Rev. Lett. {\bf 68}, 1383 (1992).
\bibitem{LuhmanWest_PRL2008}
D. R. Luhman, W. Pan, D. C. Tsui, L. N. Pfeiffer, K. W. Baldwin, and K. W. West
Phys. Rev. Lett. {\bf 101}, 266804 (2008)
\bibitem{BaoPRL2010}
W. Bao {\em et al.}, Phys. Rev. Lett. {\bf 105}, 246601 (2010).
\bibitem{Yacobi2009}
B. E. Feldman, J. Martin, and A. Yacobi, Nat. Phys. {\bf 5}, 889 (2009). 
\bibitem{WeitzYacobi2010} 
R. T. Weitz, M. T. Allen, B. E. Feldman, J. Martin, and A. Yacoby, Science {\bf 330}, 812 (2010).
\bibitem{MayorovNovoselovScience2011}
A. S. Mayorov {\em et al.}, 
Science {\bf 333}, 860 (2011). 
\bibitem{Velasco2012}
J. Velasco Jr. {\em et al.},  
Nat. Nanotech. {\bf 7}, 156 (2012).
\bibitem{ketterle_zwierlein_imbalanced_SF} M.~W.~Zwierlein, A.~Schirotzek, C.~H.~Schunck, and W.~Ketterle,
Science {\bf 311}, 492 (2006).
\bibitem{zwierlein_bose_fermi} J.~W.~Park, C.-H.~Wu, I.~Santiago, T.~G.~Tiecke, S.~Will, P.~Ahmadi, and M.~W.~Zwierlein,
Phys.~Rev.~A {\bf 85}, 051602(R) (2012).
\bibitem{ketterle_zwierlein_BEC_BCS} M.~W.~Zwierlein, J.~R.~Abo-Shaeer, A.~Schirotzek, C.~H.~Schunck, and W.~Ketterle,
Nature {\bf 435}, 1047 (2005).
\bibitem{jin_unitarity_dynamics} P.~Makotyn, C.~E.~Klauss, D.~L.~Goldberger, E.~A.~Cornell, and D.~S.~Jin, Nat. Phys. {\bf 10}, 116 (2014).
\bibitem{MoorePRL2006}
S. Mukerjee, C. Xu, and J. E. Moore, Phys. Rev. Lett. {\bf 97}, 120406 (2006).
\bibitem{Podolsky2009} D. Podolsky, S. Chandrasekharan, and A. Vishwanath,
Phys. Rev. B {\bf 80}, 214513 (2009).
\bibitem{olshanii} M.~Olshanii, Phys. Rev. Lett. {\bf 81}, 938 (1998).
\bibitem{petrov_shlyapnikov_PRL} D.~S.~Petrov, M.~Holzmann, and G.~V.~Shlyapnikov, Phys. Rev. Lett. {\bf 84}, 2551 (2000).
\bibitem{petrov_shlyapnikov_PRA} D.~S.~Petrov and G.~V.~Shlyapnikov, Phys. Rev. A {\bf 64}, 012706 (2001).
\bibitem{demler_pekker} For the inclusion of many-body effects see 
V.~Pietil\"a, D.~Pekker, Y.~Nishida, and E.~Demler, Phys. Rev. A {\bf 85}, 023621 (2012).
\bibitem{bloch_review} I.~Bloch, J.~Dalibard, W.~Zwerger, Rev. Mod. Phys. {\bf 80}, 885 (2008).
\bibitem{landau_QM}  L.~D.~Landau and E.~M.~Lifshitz, {\it Quantum Mechanics: Non-Relativistic Theory} (Pergamon Press, New York, 1987). 
\bibitem{demler_polaron_vs_kohl} D.~Pekker, M.~Babadi, R.~Sensarma, N.~Zinner, L.~Pollet, M.~W.~Zwierlein, and E.~Demler,
Phys.~Rev.~Lett. {\bf 106}, 050402 (2011).
\bibitem{jochim_CIR} S.~Sala {\it et al.}, Phys. Rev. Lett. {\bf 110}, 203202 (2013).
\bibitem{kohl_CIR} B.~Fr\"ohlich, M.~Feld, E.~Vogt, M.~Koschorreck, W.~Zwerger, and M.~K\"ohl, Phys. Rev. Lett. {\bf 106}, 105301 (2011).
\bibitem{Sommer2012} 
A. T. Sommer, L. W. Cheuk, M. J. H. Ku, W. S. Bakr, and M. W. Zwierlein,  Phys. Rev. Lett. {\bf 108}, 045302 (2012).
\bibitem{esslinger_CIR} H.~Moritz, T.~St\"oferle, K.~G\"unter, M.~K\"ohl, and T.~Esslinger, Phys. Rev. Lett. {\bf 94}, 210401 (2005).
\bibitem{spin_dependent_lattices1} \marton{
P.~J.~Lee, M.~Anderlini, B.~L.~Brown, J.~Sebby-Strabley, W.~D.~Phillips, and J.~V.~Porto,
Phys. Rev. Lett. {\bf 99}, 020402 (2007).}
\bibitem{spin_dependent_lattices2} \marton{
O.~Mandel, M.~Greiner, A.~Widera, T.~Rom, T.~W.~H\"ansch, and I.~Bloch,
Phys. Rev. Lett. {\bf 91}, 010407 (2003).}
\bibitem{spin_dependent_lattices3} \marton{
P. Soltan-Panahi \textit{et al.}, Nat. Phys. {\bf 7}, 434 (2011).}
\bibitem{footnote0} \marton{
In particular, we use the pseudopotential
$V_{\alpha\beta}(\br) \, \bigl( \dots \bigr) = 
\left( 4\pi \hbar^2 a_{\alpha\beta}/m\right) \, \delta(\br) \, \frac{\partial}{\partial_r} \bigl( r \, \dots  \bigr)$
which is a good approximation} 
whenever the trapping potential is approximately constant within the effective 
range of interactions, $R_e$~\cite{petrov_shlyapnikov_PRA, bloch_review}. 
Then, $R_e$ is required to be small compared to the oscillator length, $l_z \equiv \sqrt{\hbar/(m\omega_z)}$,
and layer separations are also restricted to the range $d_{\upa\downa} \ll l_z^2 / R_e$.
Since $R_e \sim 1-10\,\mathrm{nm}$ for most atoms~\cite{petrov_shlyapnikov_PRA, Castin}, 
these conditions are very well satisfied in standard experiments.
For the case of extremely strong trapping frequencies, where
$l_z \sim R_e$, see Refs.~\onlinecite{tight_confinement_1} and \onlinecite{tight_confinement_2}.
\bibitem{footnote_QB_molecule} \martonnew{
Throughout this paper, we refer to the scattering states 
of resonant scattering amplitudes as quasi-bound molecular states. 
The width of these resonances are identified as the inverse lifetime of
the quasi-bound state.
}
\bibitem{WickeDruten2011}
P. Wicke, S. Whitlock, and N. J. van Druten,
arXiv:1010.4545.
\bibitem{Kinoshita2004} T. Kinoshita, T. Wenger, and D. S. Weiss, Science {\bf  305}, 1125 (2004). 
\bibitem{Paredes2004} 
B. Paredes {\it et al.}, 
Nature {\bf 429}, 277 (2004).
\bibitem{Fu_paper} \marton{W.~Fu, Z.~Yu and X.~Cui, Phys. Rev. A {\bf 85}, 012703 (2012).}
\bibitem{footnote__fermions_vs_bosons} 
Although many-body effects are different for  and fermionic gases,
the two-particle scattering properties, discussed in this Section are the same for bosons and 
fermions with the modification that fermionic atoms are non-interacting in the 
s-wave channel~\cite{bloch_review}, implying $a_{\upa\upa}=a_{\downa\downa}=0$.
\bibitem{footnote_Hrel} \marton{For notational simplicity, we omit the indices $\alpha \beta$ of
$\mathcal{H}_\mathrm{rel}$ throughout the paper.}
\bibitem{footnote__symmetry_for_bosons_and_fermions}
The relative wave function $\Psi_\mathrm{rel}$  must be symmetrical 
for bosons, and antisymmetrical for fermions.
\bibitem{footnote2} Throughout this paper we assume that a set of real harmonic oscillator wave functions is used.
\bibitem{footnote__factor_of_two} 
In case of identical bosons, $\alpha=\beta$, the right hand side of Eq.~\eqref{eq:cross_section} 
contains an extra factor of two~\cite{petrov_shlyapnikov_PRA}.
\bibitem{messiah} A. Messiah, \textit{Quantum Mechanics, Volume II} (Elsevier Science B. V., 1961).
\bibitem{footnote__exact_scatt_state} 
This wave function is essentially exact at separations larger than the 
range of atom-atom interaction.
\bibitem{footnote3} We speculate that these bound states, shifted to positive energies,
may create Feshbach resonances in the interaction both in interlayer and in interlayer scattering,
if some kind of anharmonicity resonantly couples the relative and center of mass motion of particles.
Such a coupling may arise from a difference between the trapping frequencies of the two layers
(e.g. in case of bosonic mixtures), or from anharmonicities of the trapping potential~\cite{jochim_CIR}.
\bibitem{footnote_large_separations} For standard trapping frequencies 
$\omega_z/(2\pi) = 1-100 \, \mathrm{kHz}$, depending also on the atomic species
used in the experiment, our approximations are valid up to
rather large layer separations, $d_{\upa\downa}/l_z \sim 3-10$~\cite{footnote0}.
\bibitem{footnote5} Indeed, the excited states transformed into the basis of center of mass and relative coordinates are given by 
$\ket{n_\alpha=1,n_\beta=0} = (\ket{N=1,\nu=0} + \ket{N=0, \nu=1})/\sqrt{2}$.
\bibitem{footnote4} Notice that  Fig.~\ref{fig:QB}  displays only the relative motion's energy, and to obtain 
the total energy, one should add the center of mass motion's energy, $N \hbar\omega_z + Q^2/4m$.
\bibitem{footnote6} Although in case of extreme trapping frequencies, $\omega_z / 2\pi \gg 100~\mathrm{kHz}$,
the pseudopotential, describing the pair interactions between the particles, may require energy dependent 
corrections~\cite{tight_confinement_1,tight_confinement_2}, we expect the overall scattering behavior in the confined  system to be only slightly modified by these terms.
\bibitem{footnote__experimental_parameters} 
The magnetic field gradients required in this case 
to reach a separation $d_{\upa\downa} \sim l_z$ are
rather small, $\nabla B_z \sim 0.01-1~\mathrm{G/mm}$, 
they nevertheless lead to a weak magnetic field 
difference between the centers of the layers, 
in the range of $\Delta B \sim 10-100~\mathrm{mG}$. 
In case of very narrow Feshbach resonances whose width are comparable to $\Delta B$, 
this can result in a spatially varying value of the scattering lengths, 
making our results inapplicable in this case.
\bibitem{Rb_interspecies_resonance} A.~Widera {\it et al.}, Phys. Rev. Lett. {\bf 100}, 140401 (2008).
\bibitem{chin_review} C.~Chin, R.~Grimm, P.~Julienne, and E.~Tiesinga, Rev. Mod. Phys. {\bf 82}, 1225 (2010).
\bibitem{bosonic_mixtures_1} G. Modugno, M. Modugno, F. Riboli, G. Roati, and M. Inguscio, Phys. Rev. Lett. {\bf 89},
190404 (2002).
\bibitem{bosonic_mixtures_2} G. Thalhammer, G. Barontini, L. De Sarlo, J. Catani, F. Minardi, and M. Inguscio, 
Phys. Rev. Lett. {\bf 100}, 210402 (2008).
\bibitem{bosonic_mixtures_3} J. Catani, L. De Sarlo, G. Barontini, F. Minardi, and M. Inguscio, Phys. Rev. A {\bf 77},
011603(R) (2008).
\bibitem{bosonic_mixtures_4} S. B. Papp, J. M. Pino, and C. E. Wieman , Phys. Rev. Lett. {\bf 101}, 040402 (2008).
\bibitem{footnote7} The results of this paper can be carried through to bosonic mixtures with different masses, 
$m_{\upa}$ and $m_{\downa}$, but trapped with equal trapping frequencies, $\omega_z$. By introducing the center of mass, 
$M_{\alpha\beta} = m_\alpha + m_\beta$, the reduced mass,
$\mu^r_{\alpha\beta} = \frac{m_\alpha m_\beta}{m_\alpha + m_\beta}$, and the oscillator length,
$l^z_{\alpha\beta} = \sqrt{\hbar/(2 \mu^r_{\alpha\beta} \omega_z)}$, in channels $\alpha\beta$,
our results can be simply rewritten by appropriately substituting
these values for the center of mass $2m \to M_{\alpha\beta}$, the reduced mass $m/2 \to \mu^r_{\alpha\beta}$ 
and the oscillator length $l_z \to l^z_{\alpha\beta}$.
\bibitem{feynman_statistical_mechanics} R.~P.~Feynman, {\it Statistical Mechanics: A Set Of Lectures} (Westview Press, 1998).
\bibitem{Castin} Y. Castin in \textit{'Coherent atomic matter waves', Lecture Notes of Les Houches Summer School}, 
edited by R.~Kaiser, C.~Westbrook, and F.~David (EDP Sciences and Springer-Verlag, 2001). 
\bibitem{tight_confinement_1} D.~Blume and C.~H.~Greene, Phys. Rev. A {\bf 65}, 043613 (2002).
\bibitem{tight_confinement_2} P.~Naidon, E.~Tiesinga, W.~F.~Mitchell and P.~S.~Julienne, New. J. Phys. {\bf 9}, 19 (2007).
\bibliography{apssamp}
\end{thebibliography}
\end{document}